\documentclass[aps,pre,reprint,superscriptaddress]{revtex4-2}

\usepackage{amsmath,amssymb,graphicx}
\usepackage{enumitem}
\usepackage{hyperref}
\hypersetup{colorlinks=true, linkcolor=red, citecolor=blue, urlcolor=blue}
\usepackage{amsbsy}
\usepackage{latexsym}
\usepackage{color}
\usepackage{graphicx}
\usepackage{psfrag}
\usepackage[normalem]{ulem}
\usepackage{bm}
\usepackage{lipsum}
\usepackage{color}
\usepackage{tikz}
\usepackage{float}
\usepackage{multirow}
\usepackage{placeins}

\newcommand{\be}{\begin{equation}}
\newcommand{\ee}{\end{equation}}
\newcommand{\bea}{\begin{eqnarray}}
\newcommand{\eea}{\end{eqnarray}}

\newcommand{\comment}[1]{}

\begin{document}

\title{Glassiness and dynamic arrest in magnetic and non-magnetic colloids}

\author{Anuj Kumar Singh}
\email{anuj0630@gmail.com}
\affiliation{Department of Physics, Indian Institute of Technology, Hauz Khas, New Delhi 110016, India.}

\author{Lambert M\"unster}
\email{lambert.muenster@physik.tu-chemnitz.de}
\affiliation{Institute of Physics, Chemnitz University of Technology, 09107 Chemnitz, Germany.}

\author{Martin Weigel}
\email{martin.weigel@physik.tu-chemnitz.de}
\affiliation{Institute of Physics, Chemnitz University of Technology, 09107 Chemnitz, Germany.}

\author{Varsha Banerjee}
\email{varsha@physics.iitd.ac.in}
\affiliation{Department of Physics, Indian Institute of Technology, Hauz Khas, New Delhi 110016, India.}

\begin{abstract}
  We investigate and compare a range of indicators for glassiness in monodisperse
  magnetic and non-magnetic soft-sphere fluids at low temperatures with a view to
  exploring the effect of the magnetic moment.  We perform extensive molecular
  dynamics simulations using the Stockmayer model for magnetic fluids and a pure
  Lennard-Jones interaction for the non-magnetic case. Our investigations involve
  quenching experiments, in which both systems are rapidly cooled deep below their
  freezing temperatures. Although the Lennard-Jones fluid forms compact aggregates,
  the inclusion of dipolar interactions promotes the development of branched and open
  morphologies. After characterizing the static properties of the frozen structures,
  we focus on their dynamics. A key observable is the self-part of the van Hove
  function, which measures the probability that a particle is displaced by a distance
  $\Delta$ over time $t$. In both fluids, this function exhibits non-Gaussian
  behavior --- thereby providing a signature of dynamic heterogeneity and
  glassiness. This behavior stems from a separation of time scales between two
  distinct processes: mobile particles that escape their environments and immobile
  particles that vibrate within cages. In particular, we find a heavier tail in the
  van Hove function for the Stockmayer fluid, which is a consequence of the strongly
  correlated motion in chain-like structures found there. These findings shed light
  on core relaxation mechanisms in magnetic fluids, advancing our understanding of
  magnetically responsive colloidal systems.
\end{abstract}

\maketitle

\section{Introduction}

Glasses form an unusual, but at the same time surprisingly common, low-temperature
state of matter that is nominally metastable as compared to the crystalline
ground-state configurations \cite{Berthier2011}. Under such metastability, one
commonly interprets states as effectively glassy if their relaxation times exceed the
typical experimental time scales --- in practice often by many orders of
magnitude. Such phases are chimeras that are rigid as regular solids, but
structurally similar to liquids without long-range positional order. The hallmark of
such glassy states is their rich dynamical behavior that includes slow evolution or
aging as well as hysteresis and memory~\cite{Arceri2022}. Particular attention has
been devoted to the study of colloidal glasses, where a suspension of big particles
in a solvent reaches a dynamically arrested state.  Despite being trapped, these
systems exhibit intriguing and sometimes useful properties, such as in gels, frozen
disordered, non-periodic networks, and slow dynamics in the living cell, to name but
a few \cite{Royall2021, Krotzky2016, Berthier2011, Bursac2005}. Developing a deeper
understanding of commonalities and differences in the relaxation processes in such
frozen morphologies of different types is a key problem of the field.

A standard model of colloidal glasses provided by a fluid of Lennard-Jones particles
has been extensively investigated using analytical and computational techniques. The
bi-disperse variant has been demonstrated to be an excellent glass former showing a
full range of glassy characteristics with dramatic dynamic changes in response to
small variations in temperature or density \cite{Hu2022, Pedersen2018, Lima2012,
  Bruns2021}. One particularly well-studied feature is the stretched-exponential
decay of correlations at late times, known as alpha-relaxation~\cite{Williams1970}.
The distribution of single-particle displacements, described by the van Hove
function, emerges as a hallmark of glassy dynamics, exhibiting a distinct
non-Gaussian profile that highlights the presence of dynamical heterogeneities
\cite{Kob1997, Rahman1964, Hurtado2007, Stariolo2006}. The mean-squared displacement
exhibits intermediate time scales characteristic of a sub-diffusive plateau regime
\cite{Marty2005, Bursac2005}. More generally, the dynamics of these systems reflect
the coexistence of (temporally varying) regions of higher and lower mobility,
generating the dynamical heterogeneity. While, at any given time, the bulk of the
particles are immobile and confined within cages, the glassy features are connected
to the mobile particles which exhibit quasi-instantaneous jumps that play a crucial
role in relaxation processes. Their larger displacements extend the distribution
tail, following a universal exponential decay \cite{Chaudhuri2007}.

Most of the numerical work on colloidal glasses has focused on particles with
isotropic, short-range interactions, in particular the Lennard-Jones
fluid~\cite{Paci2005, Shneidman1998, Mausbach2020}. Other relevant systems such as
magnetic fluids consisting of particles with a significant dipole moment have largely
remained unexplored. Magnetic fluids are well described by the Stockmayer model which
incorporates long-ranged anisotropic dipole-dipole interactions along with the usual
short-range isotropic Lennard-Jones potential~\cite{Stockmayer1941}. The typical
structures formed in such systems consist of chains of particles where the dipole
moment is oriented along the chain, thus minimizing the dipolar contribution to the
energy. The equilibrium behavior of this system has been studied in some detail,
leading to an understanding of the phase
diagram~\cite{Stevens1995,Ganzenmueller2007,Ouyang2011}. In contrast the dynamical
behavior of such systems has received much less attention. The chain-forming tendency
of the particles introduces branching within the amorphous aggregates that
drastically slows down the dynamics and leads to a modification of the relaxation as
compared to non-magnetic fluids. The combination of positional disorder and the
orientational frustration of the dipoles makes for an interesting hybrid between
structural and spin glassy features~\cite{binder:86a}.  In recent work, two of us
studied the self-assembly of Stockmayer particles after performing quenches from the
high-temperature homogeneous phase to lower temperatures into the region of phase
coexistence~\cite{Singh2023accelerated, Singh2023phase}. We observed that the
condensates (droplets) exhibited density-dependent shapes composed of closed-packed
dipole chains that imparted characteristic magnetic properties. At even lower
temperatures, we observed a pronounced freezing effect alike to the glassy phase in
the Lennard-Jones system. This state is the subject of the current work.

In the present paper, we undertake a comprehensive study of the glassiness that
arises in amorphous, low-temperature Stockmayer aggregates with a locally ordered
morphology that is distinct from conventional structural glasses. There are several
questions that must be understood in the presence of magnetic inclusions: How do
dipolar chains affect the relaxation dynamics in frozen magnetic fluids? How do
(spatial) disorder and spin frustration respond to each other? Do the magnetic
moments lead to additional relaxation mechanisms as compared to the non-magnetic LJ
systems? To what extent do dipole-dipole interactions influence the dynamical
heterogeneity? The present study explores some of these consequential questions by
performing deep quenches of the Stockmayer as well as the Lennard-Jones fluid much
below the freezing temperature, to elucidate the role of positional disorder and
dipolar moments in complex relaxation processes. Using molecular dynamics
simulations, our comprehensive analysis reveals that these rapid quenches drive the
fluids through a gas-liquid phase transition, resulting in the formation of
kinetically arrested amorphous aggregates. The key findings of our comparative study
are as follows: (i) SM aggregates are branched because of the formation of local
domains arising from co-aligned dipole chains. However, they do not exhibit
long-range magnetic order. The non-magnetic Lennard-Jones aggregates, on the other
hand, are compact. (ii) In both fluids, the innermost particles are confined in cages
formed by the neighboring particles, whereas the surface particles experience a
softer environment and have a higher possibility of escape. Thus, the relaxation
dynamics predominantly originates from the surfaces of the frozen aggregates.  (iii)
The evaluation of the van Hove function in both fluids reveals a non-Gaussian
profile, which demonstrates the {\it dynamical heterogeneity}: The particles confined
to cages exhibit only harmonic vibrations and hence are quasi-static; the particles
in the periphery undergo substantive displacements.  (iv) Smaller displacements in
both Lennard-Jones and Stockmayer fluids exhibit a Gaussian distribution. The larger
displacements corresponding to both fluids follow heavy tails which, however, we find
to be best represented by power-law distributions instead of exponential tails. The
tails for the magnetic fluid are significantly more pronounced than for the
Lennard-Jones system.

The rest of this paper is organized as follows: In Sec.~\ref{sec2} we outline the
model and techniques employed to describe the static and dynamic properties of
colloidal glasses. Section~\ref{sec3} delves into the simulation specifics and
presents detailed numerical findings. Finally, Sec.~\ref{sec4} concludes with a
comprehensive summary and discussion of our findings.

\section{Theoretical framework}
\label{sec2}

\subsection{Models}

The Stockmayer model (SM) represents a fluid of magnetic colloids, i.e., idealized
spherical particles with embedded dipole moments~\cite{Stockmayer1941}. As such, it
is a natural generalization of the Lennard-Jones system (LJ) to the case of magnetic
particles forming a \emph{ferrofluid}. The interaction between a pair of particles
$i$ and $j$, with magnetic (or, if relevant, electric) moments $\mu_i$ and $\mu_j$,
separated by a distance $r_{ij}$ takes the form \cite{Leeuwen1993, Stevens1995}:
\begin{eqnarray}
\label{SM}
  U(\vec{r}_{ij}, \hat{\mu}_i, \hat{\mu}_j) = 4\epsilon\sum_{i,j}\bigg[{\bigg( \frac{\sigma}{r_{ij}}\bigg)}^{12}-{\bigg( \frac{\sigma}{r_{ij}}\bigg)}^6\bigg] \nonumber\\
  +\frac{\mu_0 \mu^2}{4\pi}\sum_{i,j}\bigg[ \frac{\hat{\mu}_i\cdot\hat{\mu}_j - 3(\hat{\mu}_i\cdot\hat{r}_{ij})(\hat{\mu}_j\cdot\hat{r}_{ij})}{r_{ij}^3} \bigg].
\end{eqnarray}
Here, $\vec{\mu}=\mu \hat{\mu}$ represents the dipole moment of particle $i$ and
$\hat{\mu}$ is the dipolar unit vector. The first sum denotes the well-known
Lennard-Jones potential with the $1/r^{12}$ repulsion encoding the excluded volume
(Pauli exclusion), and the attractive $1/r^6$ contribution relating to the effective
van der Waals (London) force. The parameter $\epsilon$ encapsulates the overall strength
(depth of the energy well) of the LJ coupling, while $\sigma$ defines the equilibrium
inter-particle distance (particle diameter) $r_\mathrm{c} = 2^{1/6} \sigma$. The
second sum encodes the magnetic contribution in form of the long-range dipolar
interaction with its characteristic coupling of particle difference and dipolar
vectors. Importantly, it can be attractive, neutral or repulsive, depending on the
relative orientation of $\vec{r}_{ij}$ and $\vec{\mu}_{ij}$, and it hence carries an
intrinsic potential for frustration.  Thus, SM particles experience a combination of
isotropic short-range interactions and anisotropic long-range dipolar interactions,
which crucially shapes the overall behavior of the system.

The equilibrium behavior of the LJ and SM models is rather well understood. The
presence of the attractive $1/r^6$ term leads to the occurrence of a gas-liquid phase
transition in the LJ model, featuring a regime of phase coexistence ending in a
critical point~\cite{Frenkel2001}. The addition of magnetic moments and dipolar
interactions does not remove or fundamentally alter this behavior \cite{Smit1989,
  Van1993, Leeuwen1993, Leeuwen1994, Bartke2007, Kalyuzhnyi2007}.  The location of
the critical point, characterized by the critical temperature $T_\mathrm{c}$ and the
critical density $\rho_\mathrm{c}$, however, is strongly dependent on the size of
dipole moments $\mu$, thus altering the size and typically enlarging the coexistence
region. Specifically, increasing dipole strength leads to a significant increase in
$T_\mathrm{c}$, while $\rho_\mathrm{c}$ decreases slightly. The overall effect of
chain formation on the phase behavior of the model can be described via an extension
\cite{Dudowicz2004,Hentschke2007} of Flory-Huggins theory for lattice polymers
\cite{Flory1942, Huggins1942}.

\subsection{Methodologies} 

The amorphous condensates that are the focus of the present work have neither
magnetic nor long-range positional order. In order to characterize the structurally
disordered, glassy state we consider a number of suitable observables that describe
both the static as well as the dynamic features of the system.

\subsubsection{Magnetization and Edwards-Anderson order parameter}
\label{sec_mag}

The dipolar interaction favors the formation of chains of magnetic (or, if
appropriate, electric) dipoles with moment alignment along the chains. In
sufficiently dense, positionally ordered assemblies, this will lead to a significant
overall magnetization (or polarization) of the system, which we here define as
\begin{equation}
{\bf M} = \sum_{i=1}^{N} \hat{\mu}_i/N,
\label{eq:magn}
\end{equation}
where $N$ denotes the number of particles. Magnetization can range from 0 to 1,
$M=|{\bf M}|=1$ indicating perfect ferromagnetic order and $M=0$ being asymptotically
attained in disordered (paramagnetic) states as well as for antiferromagnetic and
striped configurations. However, $M$ alone is not an appropriate order parameter for
describing the local arrangements of dipole moments within frozen morphologies.
While the dipolar particles form long chains that co-align to form ordered domains,
at lower temperatures we observe smaller and randomly oriented domains due to the
freezing of the dynamics including the orientational degrees of freedom. An
appropriate order parameter for capturing the arrangements of dipolar particles
inside such morphologies is the Edwards-Anderson (EA) parameter defined as
\cite{Parisi1983, binder:86a}:
\begin{equation}
  q_\mathrm{EA}=\left[\frac{1}{N} \sum_{i=1}^N   |\langle \hat{\mu}_i \rangle_t| ^2\right]_\mathrm{av},
  \label{qEA}
\end{equation}
where $\langle \hat{\mu}_i\rangle_t$ denotes the thermal or dynamic average of the
$i^{\mathrm{th}}$ particle according to
\begin{equation}
   \langle \hat{\mu}_i \rangle_t = \frac{1}{N_t} \sum_{t=1}^{N_t} \hat{\mu}_i(t),
\end{equation}
which is non-zero for frozen dipolar particles. Here, $[...]_\mathrm{av}$ is an
ensemble average. In the paramagnetic phase, $q_\mathrm{EA}=0$ together with
$M=0$. In the ferromagnetic phase, both $q_\mathrm{EA}\ne0$ and $M\ne0$. In the
frozen (glassy) phase, on the other hand, $q_\mathrm{EA}\ne0$ but $M\simeq0$.

\subsubsection{Pair correlation function}
\label{sec_pcf}

The degree of positional order is well characterized by the pair correlation function
(PCF, also known as the radial distribution function) as a standard probe of the
arrangement of particles within the condensate. It measures the density of particles
as a function of distance $r$ from a reference molecule. It is usually normalized by
the corresponding function found for the ideal gas, such that
$g(r) = \langle \overline{\rho(r)}\rangle/\rho_0$, where $\rho_0=N/V$ is the density
of the ideal gas and $\overline{\rho(r)}$ is the average density of the system around
$r$. The numerical evaluation is facilitated by the following formula
\cite{Weis1993,Frenkel2001}:
\begin{equation} \label{PCF}
  g(r)=\frac{1}{N\rho_0}\bigg\langle\sum_{\stackrel{i,j=1}{i\neq j}}^N\frac{\delta(r-r_{ij})}{(4/3)\pi[(r+\Delta r)^3-r^3]}\bigg\rangle.
\end{equation}
The $\delta$ function is unity if $r_{ij}$ falls within the shell centered on $r$ and
is zero otherwise. By construction, $g(r)=1$ for an ideal gas, and any deviation
implies correlations between the particles due to inter-particle interactions. In the
liquid phase, $g(r)$ exhibits a large peak at small $r$ signifying nearest-neighbor
(NN) correlations followed by small oscillations that eventually approach unity at
large $r$ signifying the absence of correlations. The solid phase is characterized by
several sharp peaks at values of $r$ that correspond to the lattice spacing of the
corresponding crystal structures.

\subsubsection{Bond order parameters}
\label{sec_bop}

In phases with only partial or local positional order, it is essential to distinguish
particles that are either part of the crystal or belong to the liquid. This is
facilitated by the evaluation of the local bond order parameters (BOPs) $q_4$ and
$q_6$ \cite{Steinhardt1983, Errington2003, Shrivastav2021, Gasser2014}:
\begin{align}
\label{q1}
q_l(i)=\sqrt{\frac{4\pi}{2l+1}\sum_{m=-l}^{l}|\Bar{q}_{lm}(i)|^2},
\end{align}
with
\begin{align}
\label{q2}
\Bar{q}_{lm}(i)=\frac{1}{N_n(i)+1}\sum_{k=0}^{N_n(i)}q_{lm}(k),
\end{align}
and 
\begin{align}
\label{q3}
q_{lm}(k)=\frac{1}{N_b(k)}\sum_{j=1}^{N_b(k)}Y_{lm}(r_{kj}).
\end{align}
In Eq.~(\ref{q2}), $N_n(i)$ denotes all NNs of the particle $i$ plus the particle
$i$. In Eq.~(\ref{q3}), $N_b(i)$ denotes only the NNs of $i$. The NNs are identified
as those that lie within a cut-off distance, which is determined by the first minimum
of the pair correlation function. The functions $Y_{lm}(r_{ij})$ are the spherical
harmonics, with $l$ as a free integer parameter and $m =-l,\ldots,l$. The BOPs
$q_l(i)$ have characteristic values for different structures which are indicated in
Table~\ref{q}. As a result, the BOPs are typically used to classify the degree and
nature of locally observed order.

\begin{table}[tb!]
  \caption{Values of $q_4$ and $q_6$ for some standard lattice structures
    \cite{Steinhardt1983, Gasser2014}.}
  \begin{ruledtabular} 
    \begin{tabular}{cccccc}
      BOP &  SC & BCC  & FCC & HCP  & BCO  \\
      \hline
      $q_4$ &0.764 &0.509 & 0.190 &0.097 &0.200\\
      $q_6$ &0.354 &0.629 & 0.575 &0.484 &0.566  \\
    \end{tabular}
  \end{ruledtabular}
  \label{q}
\end{table}

\subsubsection{Softness order parameter}
\label{sec:softness}

The occurrence of more and less mobile particles is a characteristic feature of
colloidal glassy systems. These are closely connected to the properties of the local
(nearest-neighbor) environments~\cite{Sahu2024}. Faster particles have relatively
softer surroundings.  In particular, loosely packed particles with fewer neighbors
are more likely to undergo rearrangements, a characteristic feature of structural
relaxation. Connecting to these observations, a \emph{softness order parameter} has
been derived from a mean-field theory and dynamic density functional theory
\cite{Sharma2022, Piaggi2017entropy}.  It is defined as the inverse depth of the
caging potential that is calculated in a mean-field approximation assuming that the
particle undergoes short-time dynamics in a frozen background. Using the
Ramakrishnan-Youssouf free energy functional, the mean-field potential felt by the
particle has been evaluated as \cite{Nandi2021, Sharma2022, Sahu2024}:
\begin{equation}
 \beta \Phi_i(\Delta r=0) = - \rho \int {d\vec{r}}\, C_i(r) g_i(r), 
 \label{eq:softness}
\end{equation}
where $\rho$ denotes the density of the particles. Applying the hypernetted chain
(HNC) approximation \cite{Hansen2006}, the direct correlation function can be
approximated as $C_i(r)\approx g_i(r)-1$. In addition, the modified pair correlation
function $g_i(r)$ for each particle is represented as a superposition of Gaussians
\cite{Piaggi2017}:
\begin{equation}
 g_i(r) =  \frac{1}{\rho {d\vec{r}}} \sum_j \frac{1}{\sqrt{2\pi \delta^2}} e^{-}\frac{(r-r_{ij})^2}{2\delta^2},
 \label{eq:softness2}
\end{equation}
with $\delta$ characterizing the variance of the Gaussian distribution used to smooth
the correlation function. For the 3D system, ${d\vec{r}}=4\pi r^2dr$, reflecting the
radial symmetry of the pair correlation function. The softness order parameter is
then quantitatively defined as the inverse of the caging potential \cite{Sahu2024}:
\begin{equation}
 S_i = \frac{1}{\beta \Phi_i(\Delta r=0)}.
 \label{eq:softness3}
\end{equation}
$S_i$ is a measure of local stiffness based on the immediate neighborhood of the
particle. Particles at the boundaries have a less defined environment and wider cages
and correspondingly higher softness. $S_i$ can be used to identify soft and hard
environments. Hard environments encompass particles with $S_i<\bar{S}$, where
$\bar{S}$ is the mean of the distribution $P(S_i)$ versus $S_i$. For soft
environments, $S_i>\bar{S}$. We note that these considerations do not explicitly
consider the anisotropic nature of the dipolar interaction. The actual potential is
taken into account via the explicit use of the observed pair correlation function. To
fully represent the direction dependence and coupling of positional and orientational
degrees of freedom, a generalization of the softness approach could be useful, but we
did not attempt that here.

\subsubsection{Non-Gaussianity parameter}
\label{sec_hetro}

An important feature of frozen glassy systems is the {\it dynamical heterogeneity}
that emerges due to the presence of mobile and immobile particles, leading to a broad
distribution of the relative displacements $\Delta$ of particles after a significant
evolution time. In particular, these distributions often show significant deviations
from the asymptotically expected Gaussian shape. Such deviations are frequently
quantified with the help of a non-Gaussianity parameter~\cite{Rahman1964, Kob1997}.
To define it, consider the vectorial displacements
\begin{equation} \label{displacement}
 \vec{\Delta}_i(t)= \vec{r}_i(t+t_0)-\vec{r}_i(t_0).
\end{equation}
We then consider the distribution of an (arbitrary) component $\Delta_i^\mu$,
$\mu = x$, $y$, $z$ of the particle displacements and here define the non-Gaussianity
parameter as
\begin{equation} \label{eq.alpha}
 \alpha(t) = 1- \frac{3 \left\langle \Delta^2 \right\rangle^2}{\left\langle \Delta^4 \right\rangle}.
\end{equation}
Since the system box is isotropic, we can average over components $\mu$ and hence
compute the moments of $\Delta$ as
\begin{align}
\left\langle \Delta^2 \right\rangle
    = \frac{1}{3N}\sum_{i=1}^{N} \sum_{\mu} (\Delta_i^{\mu})^2, \quad
\left\langle \Delta^4 \right\rangle
    = \frac{1}{3N}\sum_{i=1}^{N} \sum_{\mu} (\Delta_i^{\mu})^4,
\end{align}
in order to improve statistics. The parameter $\alpha(t)$ measures deviations from
the Gaussian shape. A value of $\alpha(t)=0$ corresponds to a perfectly Gaussian
distribution, while higher positive values indicate an increase in dynamical
heterogeneity within the glassy system.

\subsubsection{van Hove function}
\label{sec:CTRM}

The dynamical heterogeneity in glassy systems is best understood by directly studying
the full distribution function of the single-particle displacements, which is
conventionally known as the van Hove function.  It characterizes the probability
distribution of particle displacements over time and has proven to be a fundamental
tool in the study of dynamical heterogeneities. The {\it self} part of the van Hove
function is given by \cite{Van1954}:
\begin{equation} \label{eq.van}
  G_s(\vec{\Delta}, t) = \left< \frac{1}{N} \sum_{i=1}^N \delta\left(\vec{\Delta} + \vec{r}_i(t_0) - \vec{r}_i(t+t_0)\right)\right>,
\end{equation}
where $\vec{\Delta}$ is the relative displacement vector as defined in
Eq.~\eqref{displacement}. In practice, we study the components $G_s(\Delta^\mu,t)$
and we note that $G_s(\Delta^x,t) = G_s(\Delta^y,t) = G_s(\Delta^z,t)$. In the
following, we hence write $\Delta^\mu$ for an arbitrary component, and it is
understood that expressions can be averaged over spatial directions to improve
statistics.

Several descriptions of the behavior of $G_s$ have been proposed. One is connected to
the continuous-time random walk (CTRW) model that allows to separate out localized
vibrational motions and discrete jumps that contribute to glassy dynamics
\cite{Chaudhuri2007, Montroll1965, Monthus1996}. It starts from considering  increments
\begin{equation} \label{eq:stepsize}
  \vec{\delta}_i= \vec{r}_{i}(t+\Delta t)-\vec{r}_{i}(t),  
\end{equation}
where $\Delta t$ is a suitably chosen time step. Like for the displacements
$\vec{\Delta}$, we focus on an arbitrary component $\delta_i^\mu$ (for an isotropic
system all components are equivalent). The vibrations are characterized by a Gaussian
distribution: %
\begin{equation}
  f_\mathrm{vib}(\delta^\mu) =  (2\pi l^2)^{-3/2}\exp\Big{(}-\frac{\left[\delta^\mu\right]^2}{2 l^2}\Big{)},
  \label{eq:vibrations}
\end{equation}
with {\it zero} mean and variance $l^2$. It captures the vibrational motion within
the confined cage formed by neighboring particles. Moreover, jumps representing
escape events from these cages are also assumed to follow a Gaussian distribution,
with variance $d^2$:
\begin{equation}
  f_\mathrm{jump}(\delta^\mu) = (2\pi d^2)^{-3/2}\exp\Big{(}-\frac{\left[\delta^\mu\right]^2}{2d^2}\Big{)}.
\end{equation}

The complete formulation of the van Hove function within the framework of the CTRW
model also requires information regarding the probability $P(n,t)$ of making $n$
jumps in time $t$, and the probability $f(n, \Delta^\mu)$ of traveling a distance
$\Delta^\mu$ in $n$ jumps. The convolution of these probabilities yields
\cite{Montroll1965}:
\begin{equation}
    G_s(\Delta^\mu, t)= \sum_{n=0}^{\infty}P(n, t)f(n, \Delta^\mu).
\end{equation}
It is convenient to work in the Fourier-Laplace domain where, as we recall in
Appendix~\ref{van_Hove}, the van Hove function takes the simple form
\cite{Tunaley1974}
%
\begin{align}
    G_s(q, s)&= f_\mathrm{vib}(q)\frac{1-\phi_1(s)}{s} \\ \nonumber
    &+f(q)f_\mathrm{vib}(q)\frac{\phi_1(s)}{s}\frac{1-\phi_2(s)}{1-\phi_2(s)f(q)},
\end{align}
where $f(q) = f_\mathrm{vib}(q)f_\mathrm{jump}(q)$ is the product of the Fourier
transforms of the vibrational and jump kernels. Alongside the space distribution, one
also requires information about the distribution of waiting times. The statistical
characteristics of the time to the first jump and the intervals between consecutive
jumps are denoted by $\phi_1(t)$ and $\phi_2(t)$, respectively. It is presumed that
these distributions follow an exponential pattern, defined as follows:
\begin{equation}
    \phi_1(t) = \frac{1}{t_1}\exp\Big{(}-\frac{t}{t_1}\Big{)},
\end{equation}
\begin{equation}
    \phi_2(t) = \frac{1}{t_2}\exp\Big{(}-\frac{t}{t_2}\Big{)},
\end{equation}
where $t_1$ and $t_2$ are characteristic waiting times.  The CTRW model is thus
governed by four key fit parameters: the vibration length scale $l$, the
jump-diffusion length scale $d$, and the two characteristic time scales $t_1$ and
$t_2$. The extraction of these parameters from the simulations requires the direct
identification of jumps within the trajectories. This is typically achieved by
employing a threshold criterion in which a particle is classified as executing a jump
if the magnitude of a consecutive step exceeds a predefined cut-off $r_\mathrm{cut}$,
thereby differentiating cage breaking events from vibrational movements.

\begin{figure*}[tb!]
\centering
\includegraphics[width=0.9\textwidth]{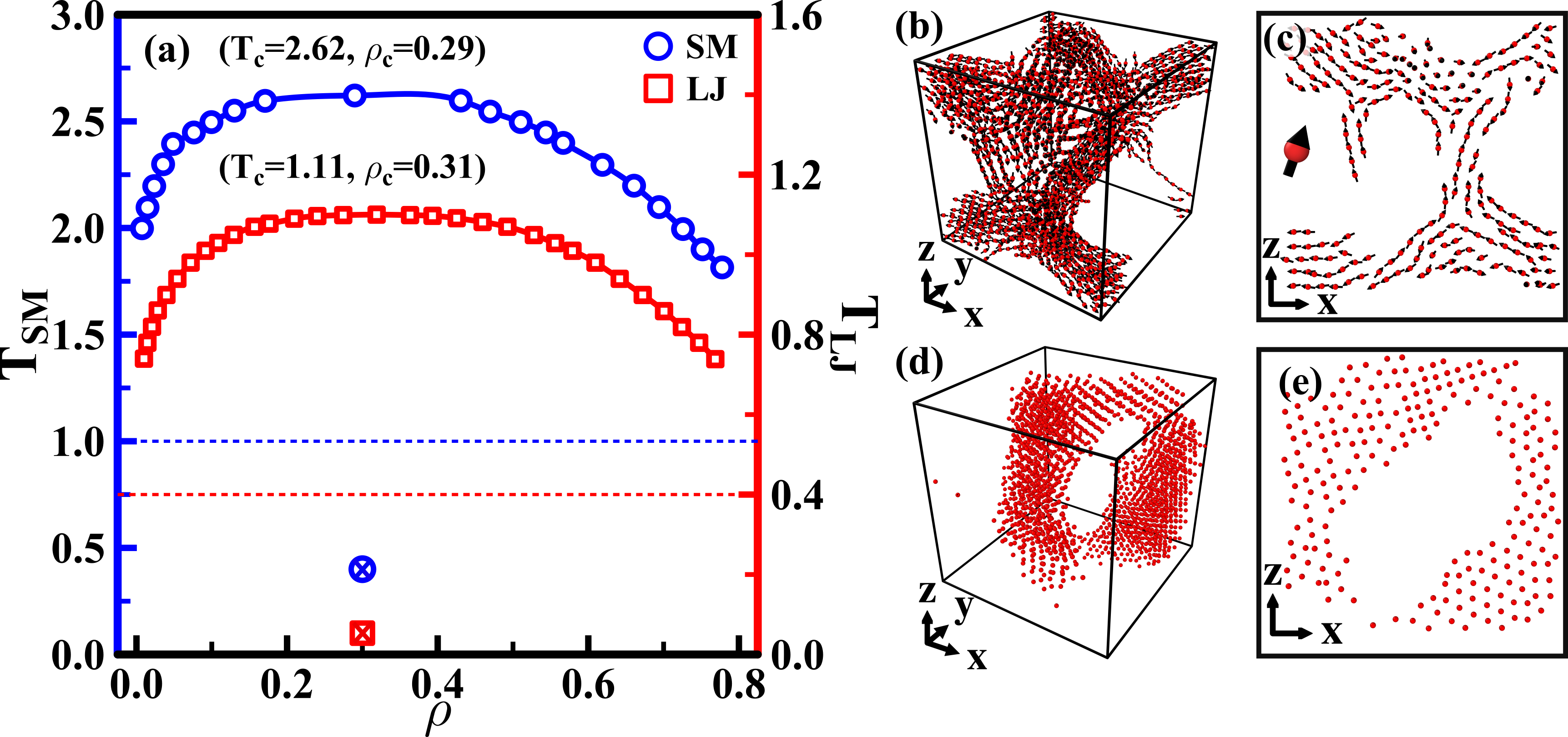}
\caption{(a) Gas-liquid coexistence diagram for the SM fluid with a dipole moment
  $\mu=2.5$ (blue, left axis) and the LJ fluid (red, right axis). The dotted lines
  qualitatively indicate the onset temperatures $T_\mathrm{f}$ of the frozen regimes
  for each system. Representative morphologies at density $\rho=0.3$ are shown for
  the SM fluid at temperature $T=0.4$ in (b) and for the LJ fluid at $T=0.1$ in
  (d). Corresponding $xz$-slices are provided in (c) and (e) to enhance
  visualization. Additionally, a prototypical SM particle is also displayed for
  clarity.}
\label{phase}
\end{figure*}

\section{Numerical simulations}
\label{sec3}

\subsection{Setup}

We performed large-scale molecular dynamics (MD) simulations for particle systems in
$d=3$ dimensions, using the canonical ($NVT$) ensemble. To allow for a systematic
comparison, we conducted separate runs with widely identical parameters (apart from
temperature and density) for the SM and LJ systems. In both cases, we applied
periodic boundary conditions using the minimum image convention \cite{Allen2017}.  We
implemented the Langevin thermostat for temperature control, incorporating both
frictional drag and stochastic forces at each particle independently alongside the
inter-particle interaction forces. The relationship between these frictional and
stochastic forces is governed by the fluctuation-dissipation theorem, which results
in temperature fluctuations around the target value and consequently maintains the
canonical ensemble for the particle systems. Because the SM system features
long-range dipolar interactions between magnetic particles, truncating the potential
could lead to significant inaccuracies. To avoid such problems, we employed Ewald
summation with metallic boundaries (i.e., excluding the surface term), which is well
suited to handle long-range interactions by virtually repeating simulation cells
\cite{Frenkel2001}.

MD simulations were conducted using the LAMMPS software package \cite{LAMMPS}. We
simulated $N$ identical non-magnetic and magnetic particles confined to a cubic box
of length $L$, with resulting density $\rho=N/V$, and interactions governed by the LJ
and SM potentials, respectively. The velocity-Verlet algorithm was used to update the
positions and velocities of the particles with a simulation time step of
$\Delta t = 0.002$. All calculations were performed in reduced LJ units, where
temperature, density, dipole moment, and time are expressed as
$ T^*= k_\mathrm{B}T/\epsilon$, $ \rho^*= N\sigma^3/V $,
$\mu^*=\mu/\sqrt{\epsilon\sigma^3}$ and
$ \Delta t^*= \Delta t/\sqrt{m\sigma^3/\epsilon} $, respectively. Here, $V$ is the
volume of the simulation box, $k_\mathrm{B}$ is the Boltzmann constant, and
$\epsilon$ and $\sigma$ are the characteristic energy and length scales,
respectively. For simplicity, we drop the asterisks in the subsequent discussion. The
LJ and SM systems were initialized in a random state and then equilibrated at a high
temperature ($T=5$, in LJ units).  The simulations were then quenched to a desired
temperature within the gas-liquid coexistence regime and allowed to evolve. After a
certain waiting period, production runs were performed, with system configurations
collected every $500$ steps. All measurements were averaged over $40$ independent
simulations.

We start with a discussion of the gas-liquid binodal curve and provide rough
estimates of the freezing lines before analyzing the kinetically trapped
low-temperature condensates. Figure~\ref{phase}(a) illustrates the coexistence
regions of LJ and SM fluids, along with their respective binodal critical
points. These systems exhibit different freezing temperatures, with
$T_\mathrm{f}\simeq 1.0$ for the SM fluid \cite{Singh2023phase} and
$T_\mathrm{f}\simeq 0.4$ for the LJ fluid \cite{Trudu2006, Khrapak2011} at
near-critical densities. Above the freezing lines, both systems exhibit a
well-defined, density-driven structural organization, as expected
\cite{Singh2023phase}. For the SM fluid (circles), following the deep quench to
$T = 0.4 < T_\mathrm{f}$, an amorphous structure is observed despite the
asymptotically expected crystalline order in a planar slab morphology at $\rho=0.3$
\cite{Singh2023phase}. A prototypical aggregate and the corresponding $xz$-slice
through the center of the system are shown in Figs.~\ref{phase}(b) and (c). The
aggregates for the SM system exhibit a locally ordered arrangement of dipole moments
and hence particles, resulting in a branched structure that contains multiple domains
with a noticeable formation of chains. For the LJ system (squares) quenched similarly
far below its own freezing line, the frozen morphology and the corresponding
$xz$-slice for the LJ fluid at $T = 0.1$ are presented in Figs.~\ref{phase}(d) and
(e), the result being consistent with the usual expectations for frozen
structures. We focus on the understanding of these commonalities and differences in
the kinetically frozen states of the two systems.

\subsection{Static properties}

We first focus on properties characterizing the static structure of the frozen states
in the SM and LJ case, in particular the immediate neighborhood of particles, their
local energy landscape (softness) as well as, for the SM system, its static magnetic
order.

\subsubsection{Magnetic order in SM condensates}

\begin{figure}[tb!]
\centering
\includegraphics[width=0.48\textwidth]{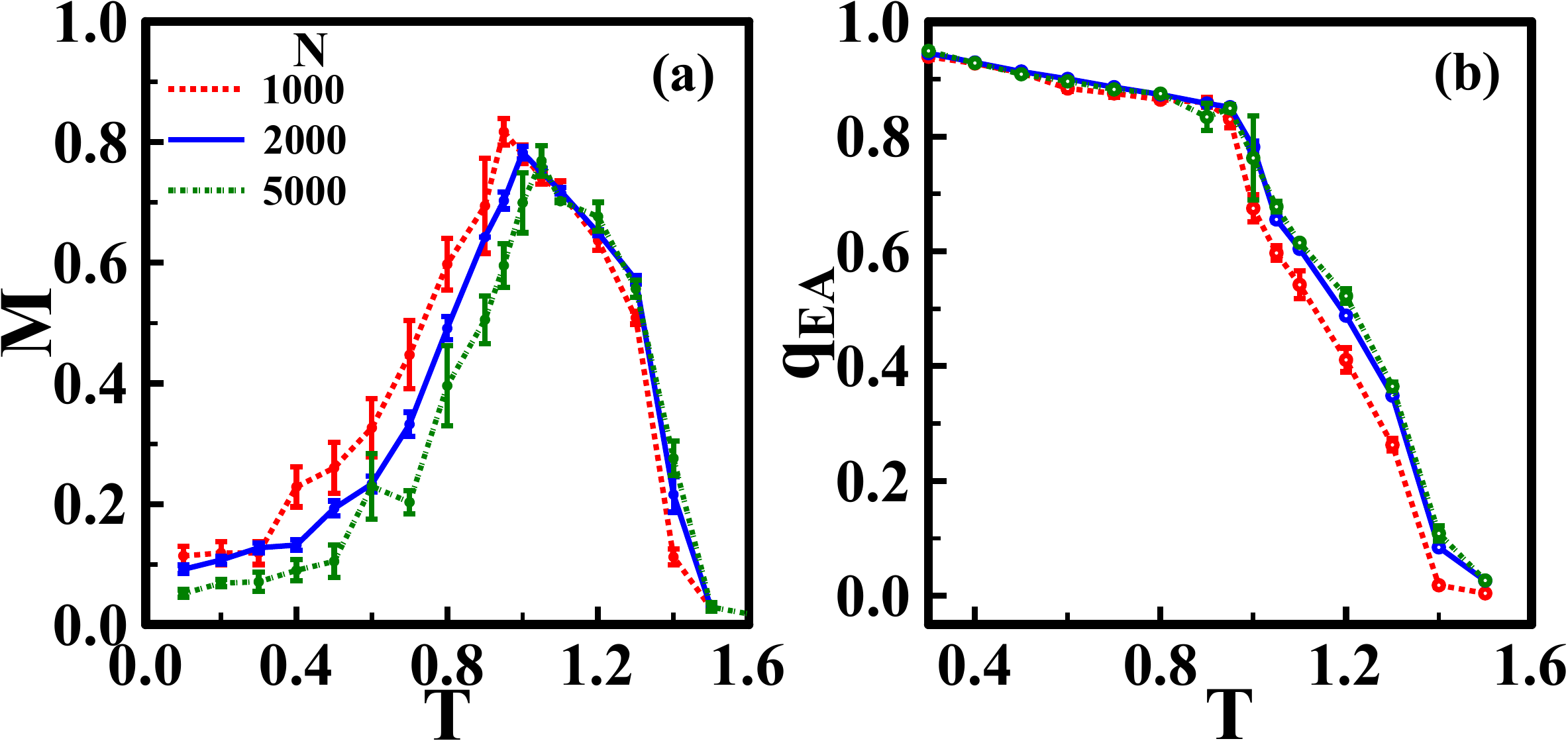}  
\caption{(a) Magnetization $M$ as a function of temperature $T$ for system sizes
  $N=1000$, $2000$, and $5000$ at a fixed density $\rho=0.3$. (b) Edwards-Anderson
  order parameter $q_\mathrm{EA}$ versus $T$ for the same simulation. Both plots
  include error bars. Data are averaged over $10$, $40$, and $5$ independent samples
  for $N=1000$, $N=2000$, and $N=5000$, respectively.}
\label{M_qEA}
\end{figure}

The SM system features dipole moments as additional degrees of freedom that are not
present in the LJ fluid. Due to the inherent coupling between relative spatial
positions and dipole orientations expressed in the dipolar potential, ordering
processes in both sectors are closely linked to each other. In Fig.~\ref{M_qEA}(a) we
show the magnetization $M$ of Eq.~\eqref{eq:magn} vs.\ temperature for different
numbers of particles, $N=1000$, $2000$, and $5000$, at a fixed density $\rho =
0.3$. The error bars shown are extracted from the fluctuations observed between
independent simulations. At higher quench temperatures, the random orientations of
the dipolar particles result in $M\simeq 0$. As the temperature is reduced, there is
alignment of the particles leading to an increase in $M$, which peaks around the
freezing temperature $T_\mathrm{f} \simeq 1.0$. The corresponding structure is a
single-domain slab with an aligned dipole, as expected at this density
\cite{Singh2023phase}. For quenches to even lower temperatures, $M$ decreases,
however, which is due to the emergence of multiple domains observed in the
prototypical morphology of Fig.~\ref{phase}(b).

While there is hence no conventional long-range order of dipole moments for
$T < T_\mathrm{f}$, the system freezes into a glassy, locally ordered state. Such
behavior is captured by the EA order parameter of Eq.~\eqref{qEA} that becomes
non-zero in systems with dynamical ergodicity breaking on the time scales of
observation. As the plot of the EA parameter as a function of temperature presented
in Fig.~\ref{M_qEA}(b) illustrates, $q_\mathrm{EA}$ increases as the freezing
temperature is approached from above and settles on values close to unity for
$T \le T_\mathrm{f}$, indicative of the frozen state. In analogy to the case of spin
glasses, where this is well established, we expect a multitude of frozen states that
are independently sampled by the system when separate runs are quenched from high
temperature~\cite{binder:86a}. Since in each of them the spins are frozen on
observational time scales, one arrives at large values of $q_\mathrm{EA}$.  In
particular, as $T \rightarrow 0$, $q_\mathrm{EA} \rightarrow 1$, signifying the
frozen state, while $M \to 0$.

\begin{figure}[tb!]
\centering
\includegraphics[width=0.48\textwidth]{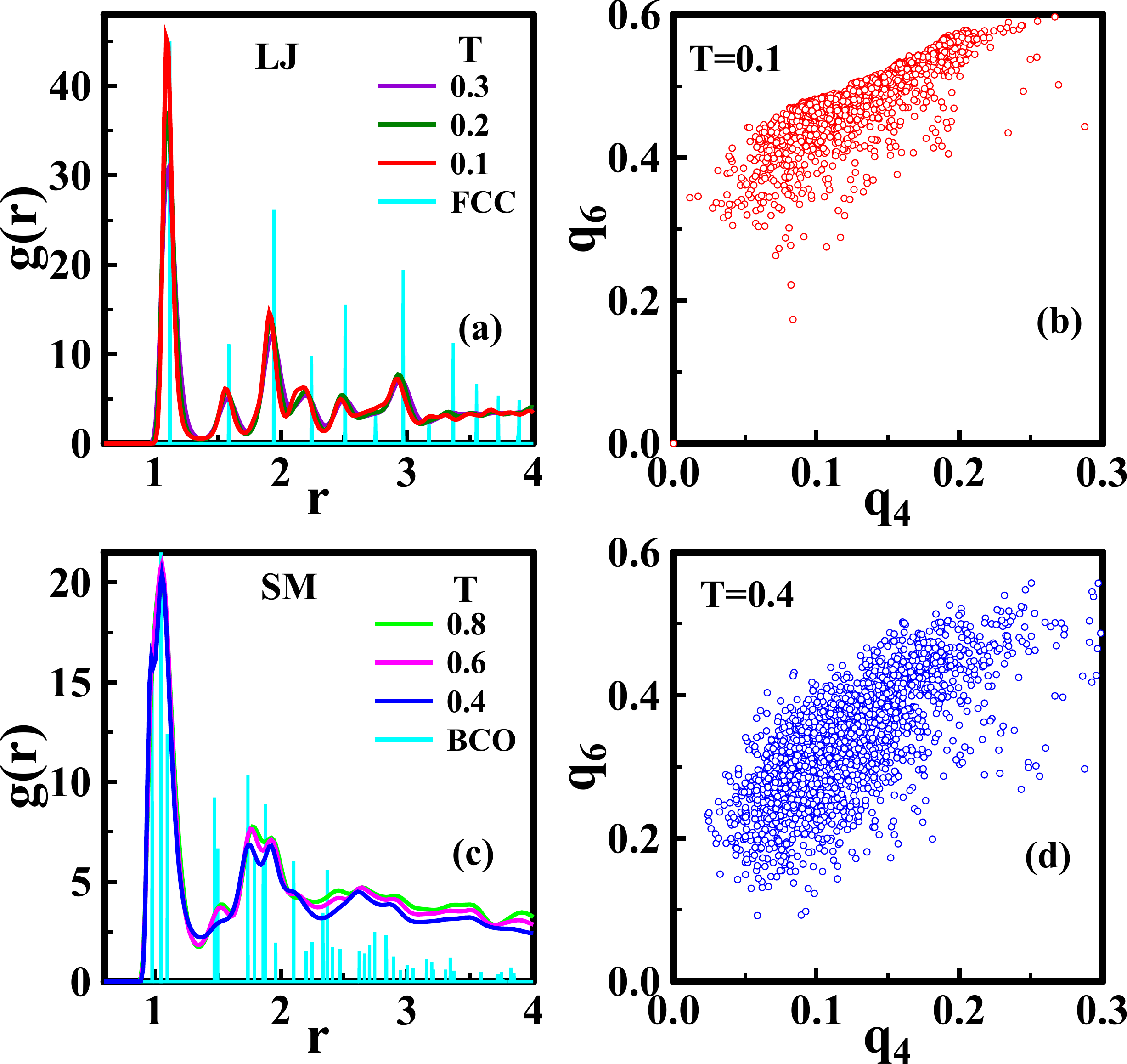}
\caption{(a) Pair correlation function $g(r)$ of the LJ fluid as a function of
  particle separation $r$ for a density $\rho=0.3$ and system size $N=2000$ at
  temperatures $T=0.3$, $0.2$ and $0.1$. For comparison, the $g(r)$ of the ideal FCC
  structure is also shown (in cyan). (b) Scatter plot of the bond order parameters
  $q_4$ and $q_6$ for $T=0.1$. (c) Similarly, $g(r)$ for the SM fluid is shown for
  the same density and system size at $T=0.8$, $0.6$, and $0.4$. The $g(r)$ of the
  ideal BCO structure is included for reference (in cyan). (d) Scatter plot of $q_4$
  versus $q_6$ for the SM case and $T=0.4$. The $g(r)$ data for both fluids are
  averaged over 100
  samples.}
\label{pcf_bop}
\end{figure}

\begin{figure*}[tb!]
\centering
\includegraphics[width=0.9\textwidth]{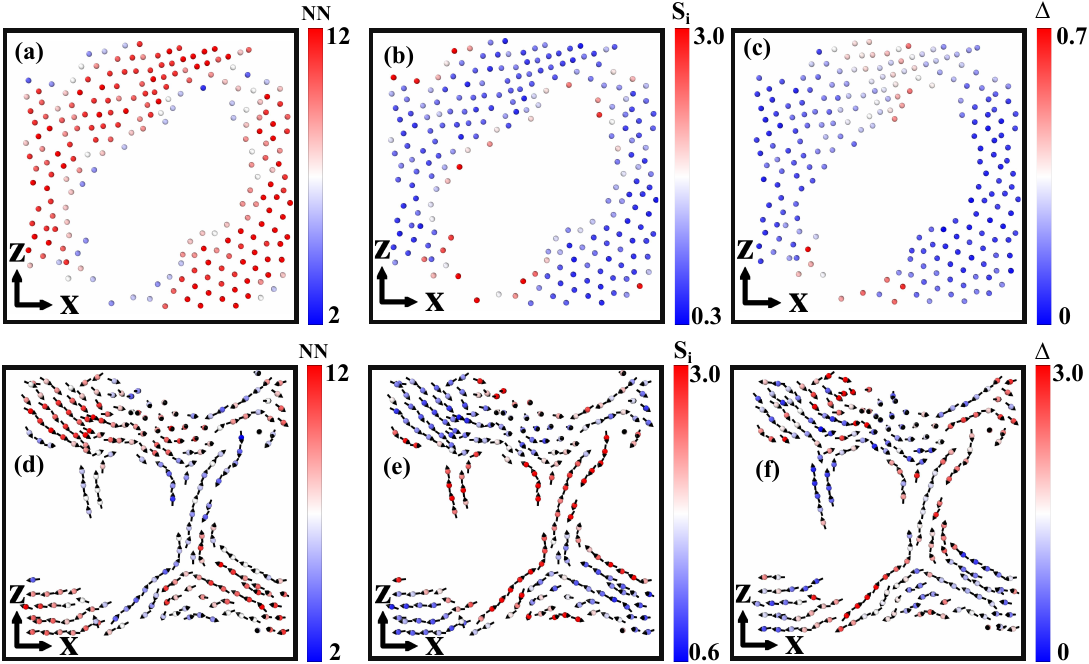}  
\caption{The top row shows prototypical $xz$-slices for the LJ fluid, corresponding
  to the configurational snapshot shown in Fig.~\ref{phase}(d). The particles are
  color-coded according to (a) the NN count, (b) the softness order parameter
  ($S^i$), and (c) the relative displacement ($\Delta$). Color bars are shown with
  each panel. The bottom row shows prototypical $xz$-slices for the SM fluid,
  corresponding to the configuration shown in Fig.~\ref{phase}(b). The particles are
  color-coded as per (d) NN, (e) $S^i$, and (f) $\Delta$, with color bars in the
  panels.}
\label{color}
\end{figure*}

\subsubsection{Deviations from crystallinity}

As previously discussed in Sec.~\ref{sec_pcf}, the PCF $g(r)$ expresses the
correlation of densities between two points separated in space by distance
$r$. Figure~\ref{pcf_bop}(a) shows $g(r)$ as a function of distance $r$ for the LJ fluid
at $T=0.3$ (purple), $0.2$ (green), and $0.1$ (red), respectively. The cyan curve
denotes the PCF of the reference face-centered cubic (FCC) structure, recognized as
the ground state for LJ or weakly dipolar fluids~\cite{Singh2025}. In this context,
the PCF for frozen morphologies seems to indicate alignment with the ground-state
structure. In this context, the local BOPs, shown in Fig.~\ref{pcf_bop}(b), serve as
a useful tool for assessing the local environment of each particle. The scatter in
the data is around $q_4\simeq 0.19$ and $q_6\simeq 0.58$, illustrating that a subset
of particles share a local environment corresponding to the FCC neighborhood, cf.\
Table~\ref{q}.

Figures~\ref{pcf_bop}(c) and \ref{pcf_bop}(d) show the results of these evaluations
for the SM fluid at $T=0.8$ (lime), $0.6$ (magenta), and $0.4$ (blue),
respectively. For comparison, the cyan curve represents the pair correlation function
of the reference body-centered orthorhombic (BCO) structure that is seen in the
ground state of highly dipolar fluids~\cite{Singh2025}. Notably, the PCF for the
frozen structures deviates significantly from the ground-state configuration,
indicating the absence of any known crystalline arrangement of particles. This
finding is further corroborated by the evaluation of the BOPs in
Fig.~\ref{pcf_bop}(d). As mentioned in Table~\ref{q}, $q_4 \simeq 0.2$ and
$q_6 \simeq 0.56$ for a BCO structure. The rather more broadly scattered data points
clearly indicate a less uniform local environment experienced by the particles, which
we mainly attribute to the observed formation of chains and the resulting
differentiation between particles in the bulk and at the surface of domains, as we
discuss next.

\begin{figure*}
\centering
\includegraphics[width=0.9\textwidth]{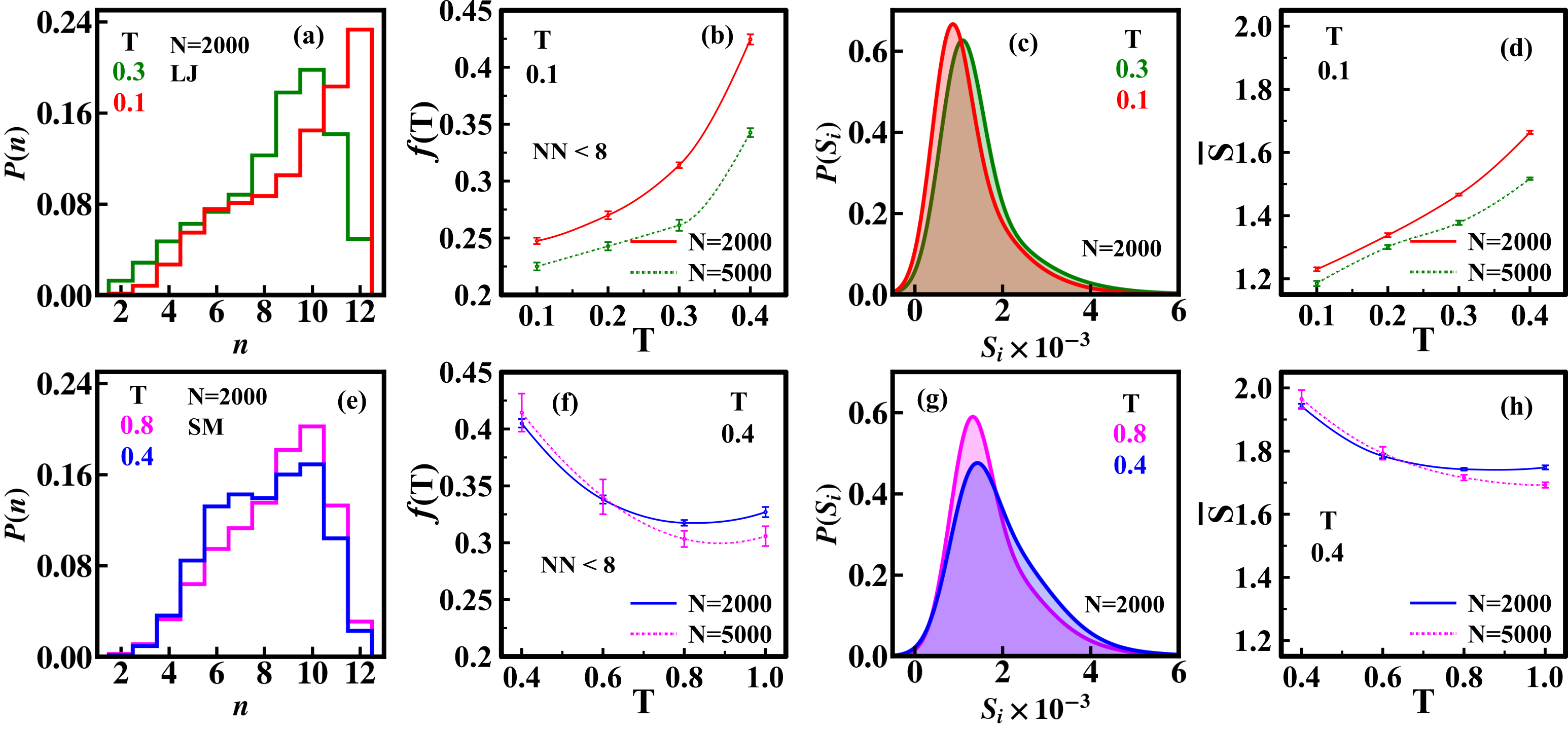} 
\caption{First row -- LJ fluid: (a) Histogram $P(n)$ of particles with $n$ NNs at
  $T=0.3$ and $T=0.1$. (b) The corresponding fraction of particles with less than
  eight NNs shown for the two temperatures. (c) The distribution of the softness
  order parameter $P(S_i)$ vs $S_i$ at $T=0.3$ and $T=0.1$. (d) Average softness as a
  function of $T$.  Second row -- SM fluid: (e) Histogram $P(n)$ vs $n$ for $T=0.8$
  and $T=0.4$. (f) Fraction of particles with less than eight NNs. (g) The
  distribution $P(S_i)$ vs $S_i$ at $T=0.8$ and $T=0.4$. (h) Average softness as a
  function of $T$. Details about the system sizes for the plotted data are provided
  in the respective
  panels.}
\label{nn_soft}
\end{figure*}

\subsubsection{Local environment and softness order parameter}

As discussed in Sec.~\ref{sec:softness}, the softness order parameter $S_i$
elucidates how tightly a particle is held by its neighbors. We evaluated it according
to Eqs.~\eqref{eq:softness}--\eqref{eq:softness3} while using $\delta = 0.02$ in
Eq.~\eqref{eq:softness2}, which we find to provide a useful compromise in terms of a
trade-off between smoothness and overfitting of noise \cite{Piaggi2017}. It serves as
a metric for predicting whether particles are likely to undergo rearrangements during
structural relaxation. This is unlikely to happen in the presence of strong caging
effects that result from complete (or nearly complete) crystallographic
environments. A relevant related proxy for the softness is hence an estimation of the
number of NNs for each particle.  In Fig.~\ref{color}(a) we present this information
for the particles in a prototypical $xz$ slice of the frozen LJ morphology shown in
Fig.~\ref{phase}(d).  Particles are color-coded on the basis of the NN count, as
indicated by the accompanying color bar. (Note that for an FCC structure, there are
12 NNs for any site.) Comparing particles in the bulk and at the surface of a domain
one hence has the (expected) behavior that the latter have fewer neighbors, on
average. The inner particles are thus less likely to escape as compared to those on
the periphery. This observation is further corroborated by the softness order
parameter $S_i$ shown in Fig.~\ref{color}(b), where tightly packed particles display
a lower softness, while loosely packed particles possess a higher softness score. In
line with these observations, Fig.~\ref{color}(c) shows the same LJ configuration,
color-coded by the (total) displacement $\Delta = |\vec{\Delta}|$ at the last
simulation time step ($t=15\times 10^3$), demonstrating that the surface particles
indeed have greater mobility relative to the interior particles. These observations
establish that the rearrangements of the particles that drive relaxation
predominantly originate from the surface of the frozen morphology.

Analogous information is obtained for the SM condensates, considering the
prototypical snapshot captured in Fig.~\ref{phase}(b). Figure~\ref{color}(d) presents
the NN information in a representative $xz$ slice. The presence of dipolar
interactions inhibits crystallization. The softness parameter of the individual
particles is shown in Fig.~\ref{color}(e). The corresponding displacements $\Delta$
of the particles are provided in Fig.~\ref{color}(f). The inference from these data
sets is that the surface particles exhibit rearrangements and the magnitude of their
movement is more pronounced compared to that of the LJ fluid. Further, they establish
that the NN count and the softness parameter are essentially equally good indicators
of particle displacement.

For further insights, we study the variation of the number of NNs and the softness
parameter as a function of temperature. Figure~\ref{nn_soft}(a) presents the
histograms of the fraction of particles with $n$ NNs, $P(n)$ vs.\ $n$ for the LJ
fluid at temperatures $T=0.3$ and $0.1$. As the quenching temperature decreases, the
fraction of particles with 12 NNs in a crystalline arrangement
increases. Figure~\ref{nn_soft}(b) shows the fraction of particles with NN $<8$,
$f(T)$ versus $T$, which is found to shrink with decreasing temperature. The softness
distribution presented in Fig.~\ref{nn_soft}(c) shifts toward smaller values at lower
temperatures, reflecting a tendency toward more compact configurations. This trend is
further highlighted in Fig.~\ref{nn_soft}(d), where the average softness shown for
two system sizes reveals the increasing dominance of crystalline particles, thus
reducing the glassiness in the LJ fluid. In contrast, the magnetic SM fluid exhibits
opposite trends for both these evaluations, see the corresponding
Figs.~\ref{nn_soft}(e)-(h). The decrease in NN and the enhanced softness observed at
lower quenching temperatures align with the emergence of a branched structure,
strongly supporting the development of glassy behavior in the SM fluid. Overall,
magnetic and non-magnetic systems exhibit opposite trends in NN count and softness
with temperature. However, in both scenarios, relaxation is initiated primarily
through surface particle rearrangements.

\subsection{Characterization of dynamical properties}

\subsubsection{Non-Gaussianity parameter}
\label{sec_hetro_}

\begin{figure*}
\centering
\includegraphics[width=0.9\textwidth]{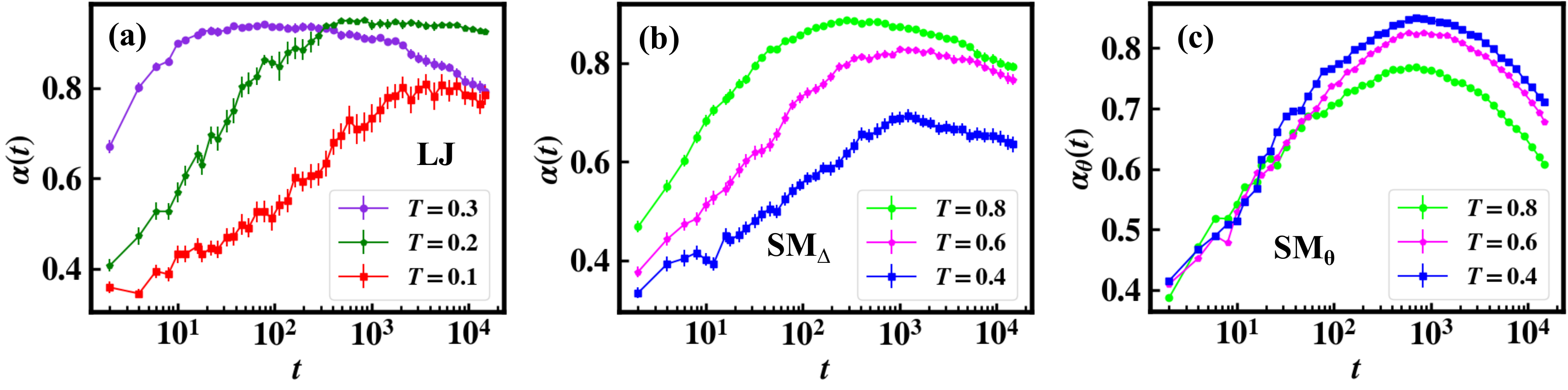} 
\caption{(a) Non-Gaussianity parameter $\alpha$ as a function of time $t$ for (a) the
  LJ fluid at $T=0.3$, $0.2$, and $0.1$ as well as (b) the SM fluid at $T=0.8$,
  $0.6$, $0.4$. (c) $\alpha(t)$ for the relative, shifted angle distribution of the
  dipole moments of the SM particles.}
\label{alpha}
\end{figure*}

In Fig.~\ref{alpha} we show the time evolution of the non-Gaussianity parameter
$\alpha(t)$ defined in Eq.~\eqref{eq.alpha}, measuring the degree of dynamical
heterogeneity in the system. In general, we see small values of $\alpha$ for early
times, representative of the ballistic motion of particles in this
range~\cite{Kob1997}. Up to intermediate times, in the regime of $\beta$ relaxation,
the values of the non-Gaussianity parameter $\alpha$ increase to deviate
significantly from $0$ which is indicative of the presence of dynamical
heterogeneities, leading to clearly non-Gaussian displacement distributions. As is
seen in panel Fig.~\ref{alpha}(a) for the LJ system, these values reach
$\alpha \approx 0.8$ and then start decaying again, as would be expected for
diffusive behavior at late times. As the temperature is lowered, we find that the
initial increase of $\alpha(t)$ is slowed, and the maximum is achieved at a later
time. In the bi-disperse LJ system it is observed that $\alpha(t)$ increases as $T$
is lowered, indicating the increase of dynamical heterogeneities in this
limit~\cite{Kob1997}. On the basis of the limited time range of our data, however, it
is not possible to decide whether that is the case here, too, or whether rather the
dynamical heterogeneities are reduced at lower temperatures due to
crystallization. The non-Gaussianity for the SM model presented in
Fig.~\ref{alpha}(b) shows a similar increase with $t$, but there is hardly any shift
of the maximum with temperature, and the heights even decrease for colder systems.
This non-trivial behavior, contrasting with conventional glassy dynamics, stems from
dipolar interactions that form extended, locally ordered aggregates. The
heterogeneous motion of these structures becomes much more coherent than the
localized cage-breaking dynamics typical of standard glass formers. Consequently,
single-particle displacements become more homogeneous (closer to Gaussian), even as
the overall structural relaxation slows down significantly and suppresses the
long-distance tail of the self-part of the van Hove function at lower $T$.

Since dipolar particles have an additional internal degree of freedom, it makes
sense to also investigate the angular displacements that capture the dipolar
reorientation events. We consider the relative angle of the dipole orientation of a
particle at different times,
\begin{equation}
  \theta_i(t) = \cos^{-1} \left(\hat{\mu}_i(t+t_0) \cdot \hat{\mu}_i(t_0)\right).
  \label{eq:relative-orientation}
\end{equation}
Since this distribution is not necessarily symmetric with respect to zero, we
consider $\alpha$ for the shifted variable
$\Delta \theta_i(t) = \theta_i(t) - \overline{\theta(t)}$, where $\overline{\theta}$ corresponds
to the distribution mean. The corresponding $\alpha$ parameter is shown in
Fig.~\ref{alpha}(c). The heterogeneity in relative orientation increases with
decreasing temperature which we attribute to the presence of additional orientational
relaxation at low temperatures. Together with dipolar interactions, this orientation
heterogeneity promotes structural fluctuations and broadens the extent of local
trapping.

\begin{figure*}[tb!]
\centering
\includegraphics[width=0.9\textwidth]{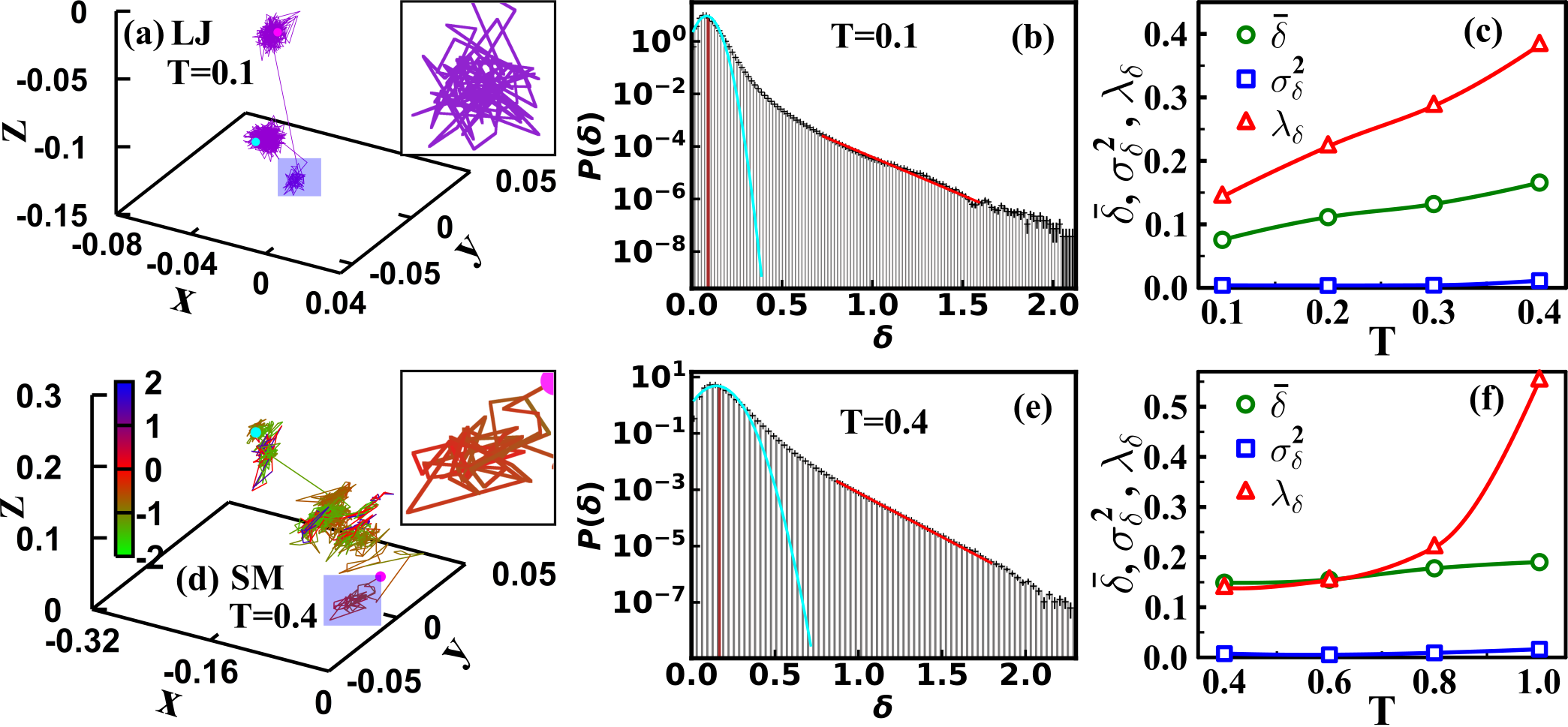}  
\caption{First row -- LJ fluid: (a) Representative trajectory of the particle that
  exhibits the maximum step length among the 2000 particles at $T=0.1$. The magenta
  and cyan markers denote the starting ($t_1=15000$) and ending ($t_2=20000$)
  points. An enlarged portion of the trajectory is shown in the inset. The
  displacement is expressed in units of box length $L$. (b) Corresponding step size
  distribution on semi-logarithmic axes for all particles ($N = 2000$), averaged over
  $100$ samples. The early data are well described by a Gaussian distribution (cyan)
  and the long tail with an exponential law (red). The vertical line indicates the
  distribution mean. (c) Variation of the mean ($\bar{\delta}$), variance
  ($\sigma_{\delta}^2$), and decay length scale $\lambda_{\delta}$ with temperature
  $T$. Second row -- SM fluid: (d) Representative trajectory of the particle with the
  maximum step length among the 2000 particles at $T=0.4$. The color scale represents
  the relative orientation of the dipole moment. The enlarged portion is included in
  an inset. (e) The corresponding step-size distribution for all particles. (f)
  Variation of $\bar{\delta}$, $\sigma_{\delta}^2$, and $\lambda_{\delta}$ with $T$.
}
\label{step}
\end{figure*}

\subsubsection{Step size distribution}
\label{sec_step}

In frozen aggregates, the majority of particles are confined within cages of varying
depths and exhibit significant vibrational motion, while -- at any given time -- a
small subset of particles are more mobile and are able to escape from these
cages~\cite{Berthier2011}. This subset contributes to rearrangement events and plays
a crucial role in understanding the relaxation mechanisms of frozen
morphologies. Here, we consider some characteristic single-particle trajectories and
the relevant distributions of step sizes. For these purposes, a time step corresponds
to a mesoscopic time unit equivalent to 1000 MD steps ($\Delta t=0.002 \times 1000=2$
in the LJ unit system). These steps are chosen such that they can capture escape
events, but it is extremely unlikely for several escape events to occur during a
single step. The step size for the $i^\mathrm{th}$ particle is defined as the
displacement between its positions at successive time steps:
\begin{equation} \label{eq:stepsize}
\delta_i= |\vec{r}_{i}(t+\Delta t)-\vec{r}_{i}(t)|,  
\end{equation}
where, as before, $\vec{r}_{i}(t)$ is the position of the $i^{\text{th}}$ particle at
time $t$.

To illustrate these features, we initially analyze the trajectory of a prominent
mobile particle chosen out of 2000 particles, at temperature $T=0.1$ in the LJ
fluid. The prototypical trajectory, shown in Fig.~\ref{step}(a), highlights the
vibrational motion confined within cages, followed by escape events. The histogram
depicting the distribution of step lengths, shown in Fig.~\ref{step}(b), reveals that
the majority of particles remain confined, with only a small fraction participating
in relaxation dynamics executing long jumps. The central part of the distribution in
Fig.~\ref{step}(b), corresponding to the vibrational motion, is fitted well by a
Gaussian distribution:
\begin{align}
    P(\delta) = \frac{1}{\sqrt{2\pi\sigma_{\delta}^2}} \exp\left(\frac{(\delta-\bar{\delta})^2}{2\sigma_{\delta}^2}\right),
  \label{Gaussian}
\end{align}
where $\bar{\delta} = 0.075554(3)$ and $\sigma_{\delta}^2 = 0.003895(1)$ are the mean
and variance of the Gaussian distribution, respectively \footnote{Note that these
  parameters result from fits to correlated data points, such that the error
  estimates can be biased.}. The fitted line (cyan) starts to deviate from this
behavior for larger values of $\delta$.  The contributions in the tail originate from
the mobile particles and can be fitted to an exponential decay law:
\begin{align} \label{exponential_fit}
    P(\delta) = b_\delta\exp\left(- \frac{\delta}{\lambda_{\delta}} \right), 
\end{align}
where $\lambda_{\delta}=0.14343(49)$ is the decay length and $b_\delta$ is a
normalization factor. The variation in the values of $\bar{\delta}$,
$\sigma_{\delta}^2$ and $\lambda_{\delta}$ as a function of temperature is shown in
Fig.~\ref{step}(c).  A decrease in quenching temperature moderately influences the
Gaussian distribution but markedly reduces the decay-length scale, such that large
escape steps become less likely at lower temperatures. This suggests that lower
temperatures effectively slow down dynamics by decreasing the particle step lengths
in frozen morphologies.

Performing an analogous analysis for the SM system, the trajectory of the particle
exhibiting the largest step length at $T=0.4$ is shown in Fig.~\ref{step}(d). In this
system, the particles show vibrational dynamics within confined cages while
occasionally undertaking long jumps, contributing to the extended tail of the
distribution. To elucidate these characteristics, the corresponding histogram of step
lengths for all particles is illustrated in Fig.~\ref{step}(e). The central part is
well fitted by a Gaussian distribution (cyan) with $\bar{\delta} = 0.148926(3)$ and
$\sigma_{\delta}^2 = 0.007533(1)$, while the histogram tail obeys an exponential
decay (red) with $\lambda_{\delta}=0.13843(21)$. The variation of the fit parameters
with temperature is reflected in Fig.~\ref{step}(f). The value $\sigma_{\delta}^{2}$
is somewhat higher in comparison to the non-magnetic system, indicating a broader
Gaussian distribution. The length scale $\lambda_{\delta}$ increases with increasing
temperature, in line with the behavior observed in the LJ fluid, but the increase
appears to be more rapid for SM as compared to LJ. Collectively, these findings
emphasize the universal relationship between temperature and particle dynamics within
frozen morphologies.

\begin{figure*}[tb!]
\centering
\includegraphics[width=0.9\textwidth]{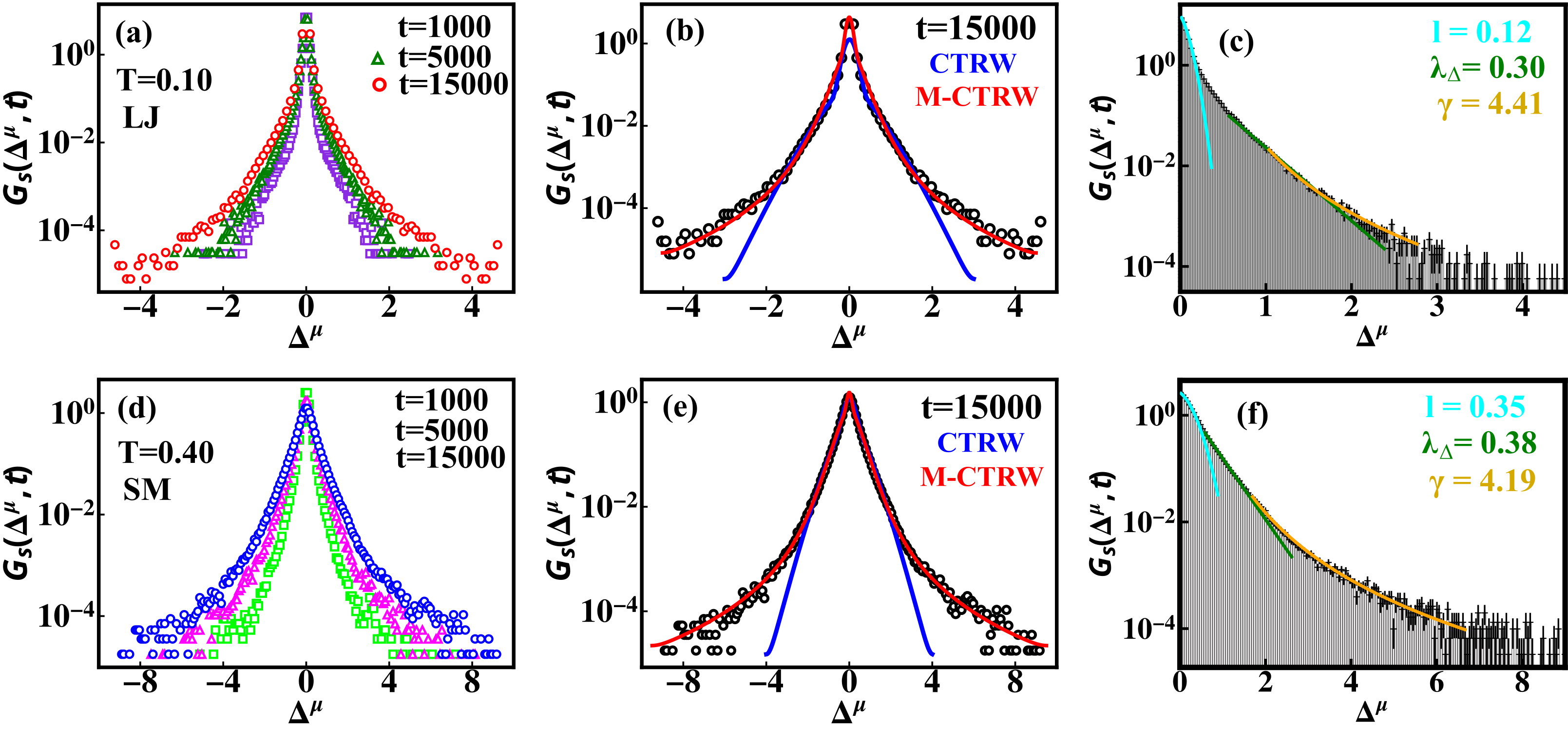} 
\caption{First row -- LJ fluid: (a) The self-part of the van Hove function
  $G_s(\Delta^{\mu},t)$ vs $\Delta^{\mu}$ is shown on a semi-logarithmic scale at
  $T = 0.1$, starting measurements after an initial waiting time $t_0 = 5000$ for
  times $t = 1000$, $5000$, and $15000$. (b) The function $G_s(\Delta^{\mu},t)$ at
  $T = 0.1$ and $t = 15000$ is shown in black, together with a fit based on the
  continuous-time random walk (CTRW) model (blue) and the modified CTRW (M-CTRW)
  model (red), defined in Appendix~\ref{M-CTRW}. (c) The histogram with positive
  folded data at $t=15000$ fits a Gaussian distribution (cyan) for short
  displacements ($\Delta^{\mu} \leq 0.2\sigma$) and with an exponential decay (green)
  and a power law decay (yellow) for larger displacements ($\Delta^{\mu} >
  0.2\sigma$). The vibrating length $l$ of Eq.~\eqref{eq:stepsize}, decay length
  scale $\lambda_{\Delta}$ of Eq.~\eqref{exponential_fit2}, and power-law exponent
  $\gamma_{\Delta}$ of Eq.~\eqref{eq:power-law} are indicated in the plot. Second row
  -- SM fluid: (d) Similarly, $G_s(\Delta^{\mu},t)$ vs $\Delta^{\mu}$ is shown on a
  semi-logarithmic scale at $T=0.4$ for $t=1000$, $5000$, and $15000$. (e)
  $G_s(\Delta^{\mu},t)$ (in black) is shown at $T=0.4$ and $t=15000$ together with
  fits of the CTRW model in blue and M-CTRM models in red. (f) The folded histogram
  at $t=15000$ is fitted with a Gaussian distribution (cyan) for small displacements
  ($\Delta^{\mu} \leq 0.2\sigma$) and with an exponential behavior (green) and
  power-law behavior (yellow) for larger displacements ($\Delta^{\mu} > 0.2\sigma$).}
\label{rel_dis}
\end{figure*}

\subsubsection{Evaluation of the van Hove function}
\label{sec_displacement}

We now turn to an investigation of the total displacements up to time $t$, i.e., the
self part \eqref{eq.van} of the van Hove function. We consider displacements
$\vec{\Delta}_i$ according to Eq.~\eqref{displacement}, focusing on an arbitrary
vector component while combining the statistics for the $x$, $y$ and $z$ components
as indicated in Sec.~\ref{sec:CTRM}. In Fig.~\ref{rel_dis}(a), we show the van Hove
function $G_s(\Delta^{\mu},t)$ for the LJ fluid at temperature $T=0.1$, measured at
times $t=1000$, $5000$, and $15000$. As the system evolves to time $t+t_0$ after an
initial waiting time $t_0$, the extended tail in $G_s(\Delta^{\mu},t)$ gradually
becomes more prominent. To comprehensively illustrate the tail of the distribution,
$G_s(\Delta^{\mu},t)$ is presented in Fig.~\ref{rel_dis}(b), alongside a fit of the
CTRW model to the data (blue curve), as defined in Sec.~\ref{sec:CTRM}. The fitting
procedure is discussed in more detail in Appendix \ref{M-CTRW}. The CTRW model
describes vibrational motions and instantaneous jumps with Gaussian distributions,
while the waiting times are modeled by an exponential decay law. This model fits the
central region of $G_s(\Delta^{\mu},t)$ well but fails to capture the tail
behavior. To improve on this shortcoming, we introduce a modified CTRW (M-CTRW) model
in which the jump statistics follow a power-law and the waiting-time distribution is
described by a generalized exponential function, see the details in
Appendix~\ref{M-CTRW}. This modified formalism accurately describes the van Hove
function over the entire range, as shown by the red fit curve. The fitting is further
emphasized by isolating vibration and jump events. Figure~\ref{rel_dis}(c) presents
the positively-folded data, distinctly separating the regimes of small and large
motions. Most particles exhibit smaller displacements ($\Delta^{\mu}\leq 0.2\sigma$),
consistent with quasi-harmonic cage vibrations; this central part is well described
by a Gaussian fit (cyan curve). In contrast, those particles exhibiting larger
displacements ($\Delta^{\mu} > 0.2\sigma$) contribute to the extended distribution
tail; they are linked to jump or escape events.  Here, we demonstrate that an
exponential fit (green curve) of the form
\begin{align} \label{exponential_fit2}
    G_s(\Delta^{\mu},t) = b_\Delta\exp\left(- \frac{\Delta^\mu}{\lambda_{\Delta}} \right), 
\end{align}
analogous to that of Eq.~\eqref{exponential_fit} for the step sizes captures
most of the data but not the long tail associated with the larger displacements of
the surface particles, whereas a power-law fit (orange curve) accurately describes this
tail. The power law fit is expressed by the following relation:
\begin{equation}
  \label{eq:power-law}
  G_s(\Delta^{\mu},t) = a_\Delta(\Delta^{\mu})^{-\gamma},
\end{equation}
where $\gamma$ represents the power-law exponent.

For comparison, the van Hove function within the magnetic SM fluid at $T=$ 0.4 is
shown in Fig.~\ref{rel_dis}(d). As the system evolves, the emergence of a long tail
becomes evident. The corresponding van Hove function, fitted using the CTRW model
(blue curve), is presented in Fig.~\ref{rel_dis}(e). The observed deviation of the
long tail from the fitted model indicates a departure from exponential behavior,
suggesting that an alternative functional form is needed to properly characterize the
tail. The M-CTRW formalism works well to capture the tail of the SM data (red
curve). In this case, a power-law form of the jumps is hence also required to fit the
van Hove function. To emphasize this further, we have isolated the smaller
displacements -- which represent the central part of the distribution -- from the
long tail of jump events, as demonstrated in the folded positive-side data in
Fig.~\ref{rel_dis}(f). Similar to the behavior observed in the LJ fluid, most
particles exhibit smaller displacements ($\Delta^{\mu}\leq 0.2\sigma$), consistent
with quasi-harmonic cage vibrations; this central region is well described by a
Gaussian fit (cyan curve). In contrast, particles that experience larger
displacements ($\Delta^{\mu} > 0.2\sigma$) are responsible for the long tail of the
distribution, which is associated with jump or escape events. First, we fitted the
tail of the distribution with an exponential function (green curve), which accounts
for only a limited subset of the data. In contrast, a power-law fit (orange curve)
captures nearly the entire tail region. The enhanced tail of the SM fluid as compared
to the LJ reference is attributed to the strongly correlated motion of the SM
particles. In general, the SM fluid exhibits significantly higher weight for particle
trajectories with a large total displacement compared to the LJ fluid.

\subsubsection{Relaxation of dipolar orientation}
\label{sec_orientation}

\begin{figure}[tb!]
\centering
\includegraphics[width=0.48\textwidth]{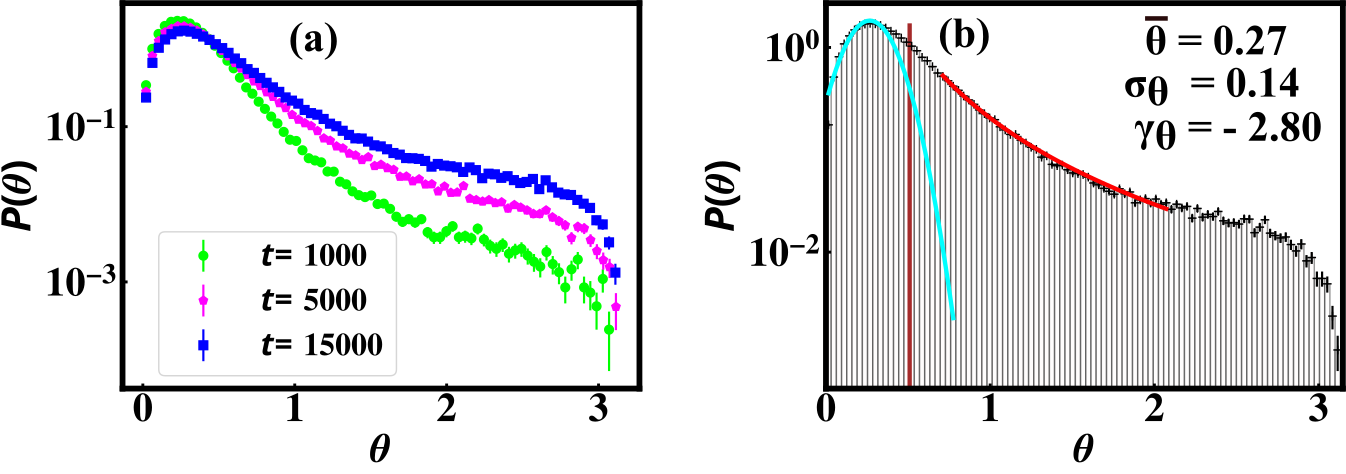} 
\caption{(a) Distribution $P(\theta)$ of the relative angular displacement $\theta$
  for SM particles for $T=0.4$ and times $t=1000$, $5000$, and $15000$. (b) The
  histogram is fitted with a Gaussian distribution (cyan) for smaller angles
  $\theta$. The long tail, attributed to escape events, is fitted with a power-law
  distribution (red). The values of the mean $\bar{\theta}$ and standard deviation
  $\sigma_{\theta}$ of the Gaussian approximation as well as the decay exponent
  $\gamma_\theta$ of the power-law tail are indicated in the plot.}
\label{rel_angle}
\end{figure}
  
Since the dipolar SM particles have an additional orientational degree of freedom, it
is interesting to investigate its role in particular for escape events. Hence, it is
also relevant to discuss the orientation dynamics as it can offer significant
insights into dynamic behavior analogous to those observed for relative
displacements. The relative orientation (or angle) of the $i^{\text{th}}$ particle at
instants $t + t_0$ and $t_0$ is given by Eq.~\eqref{eq:relative-orientation}.
Figure~\ref{rel_angle}(a) shows the distribution $P(\theta)$ vs. $\theta$ at
temperature $T=0.4$ for times $t=1000$, $5000$, and $15000$. In particular, the tail
of the distribution becomes increasingly prominent as the system undergoes temporal
evolution. To emphasize the behavior of the distribution tails, the data for
$t = 15\times 10^3$ are presented as a histogram in Fig.~\ref{rel_angle}(b), where we
distinctly separate the fits for mobile and immobile particles.  For small angular
deviations ($\theta\leq 0.5$ radians), the distribution is appropriately
characterized by a Gaussian profile, indicative of quasi-harmonic angular vibrations
within the confining cages. In contrast, large angular deviations, associated with
infrequent but substantial reorientation, give rise to extended tails in the
histogram that follow a power-law trend. These observations reveal a strong
correlation between relative orientation $\theta$ and relative displacement $\Delta$,
highlighting their interconnected roles in the regulation of the dynamics of the
glassy magnetic system.

\subsubsection{Mean square displacement}
\label{sec_msd}

\begin{figure}[tb!]
\centering
\includegraphics[width=0.48\textwidth]{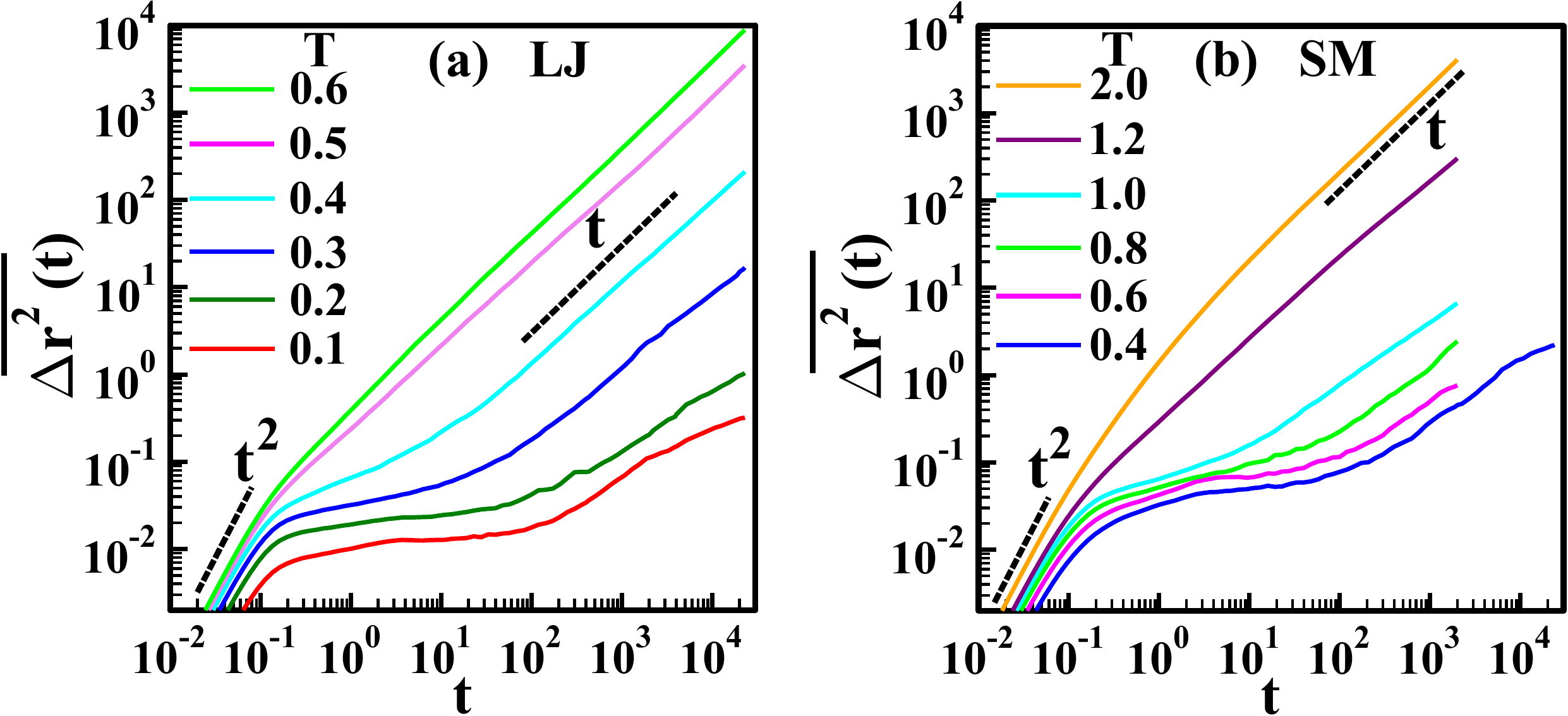}  
\caption{Variation of the MSD with time on a log-log scale at the specified temperatures
  for (a) the LJ fluid, and (b) the SM fluid. The dashed lines with specified slopes
  are guides to the eye.}
\label{MSD}
\end{figure}

To complete our discussion of particle dynamics, we analyze the mean-squared
displacement (MSD),
\[
  \Delta r^2(t) = \langle |\vec{\Delta}_i|^2\rangle =
  \langle|\vec{r}_i(t+t_0)-\vec{r}_i(t_0)|^2\rangle,
\]
where the angular brackets indicate the spatial and ensemble average. Typically,
$\Delta r^2(t) \sim t^{\gamma}$, with $\gamma=2$ at early times, indicative of
ballistic motion, and $\gamma=1$ at late times, indicating diffusive motion. In the
case of glassy dynamics, a plateau emerges at intermediate times between these two
extreme regimes. It reflects the caging effect, where particles become temporarily
trapped in the local minima of the system's energy landscape. Figure~\ref{MSD}(a)
illustrates $\Delta r^2(t)$ vs.\ $t$ of the LJ fluid at various quenching
temperatures. At high temperatures, the system exhibits both ballistic and diffusive
dynamics. As the quenching temperature decreases, the plateau becomes prominent,
indicating particle confinement within cages ($\beta$ relaxation). At sufficiently
low temperatures, the system transitions to a sub-diffusive regime for longer
simulation times.  For comparison, Fig.~\ref{MSD}(b) presents $\Delta r^2(t)$ vs.\
$t$ of the SM fluid at different temperatures. Despite variations in the critical
points and freezing temperatures of the two fluids, their MSDs exhibit similar
qualitative characteristics, however an elevated plateau is observed in the SM
fluid. This implies a wider cage, allowing the particles more space to move before
escaping and thus implying weaker dynamical confinement relative to the LJ
system. This behavior originates from anisotropic dipolar interactions that promote
branched or network-like morphologies, resulting in less compact local environments
and enhanced vibrational and orientational motion within cages.

\section{Summary and discussion}
\label{sec4}

Let us summarize our findings. We used extensive MD simulations to investigate the
low-temperature frozen structures within both magnetic and non-magnetic colloidal
systems to explore the origins of the relaxation phenomena and their role for the
dynamics of the systems. The interactions between the particles were modeled by the
LJ potential with the addition of the SM potential to represent the dipole-dipole
interaction in a magnetic fluid. For these investigations, we performed quenching
experiments from the homogeneous fluid to the coexistence regime significantly below
the respective freezing temperatures. The aggregates become kinetically arrested as a
result of caging effects. In the magnetic fluid, as a consequence of the strong
dipolar interactions, various local domains comprising dipole chains emerge,
exhibiting positional disorder alongside frustration.

We conducted a comparative analysis of positional disorder in non-magnetic and
magnetic systems. Non-magnetic fluids, as noted for large simulation times, tend to
form compact aggregates, whereas magnetic particles exhibit a branching structure,
with these tendencies becoming increasingly pronounced as the quenching temperature
decreases. In non-magnetic systems, monodisperse particles are unable to survive in a
disordered state due to the presence of shallow local barriers, subsequently
transforming towards an FCC ground state. However, in a magnetic fluid, the
chain-forming tendency of dipolar interactions intensifies local barriers, thereby
inhibiting crystallization into the ground states of BCO. In both fluids, the
interior particles are tightly bound by the neighboring molecules, whereas the
surface particles are loosely bound, participating in rearrangement during escape
events. Thus, the relaxation process originates on the surface of the observed
clusters.

From a dynamic perspective, both fluids demonstrate heterogeneity through
non-Gaussian behavior characterized by the presence (at any given time) of immobile
and mobile particles. The former exhibit quasi-harmonic vibrations within confined
cages, while the latter are involved in the relaxation processes. The van Hove
functions representing the relaxation show characteristic similarities and
differences between the two fluids. The vibrational motion in both systems follows a
Gaussian distribution. The relaxation exhibits pronounced heavy tails that are
connected to the escape events of particles from the surrounding cages. These tails
are found to be significantly more pronounced for the magnetic system that in the LJ
fluid. We attribute this enhanced tail to the occurrence of larger jumps within
branched morphologies that occur more frequently as compared to those in the
non-magnetic counterpart. The behavior of the van Hove function can be modeled using
a continuous-time random-walk model. Compared to the standard approach using Gaussian
vibration and jump distributions as well as exponential waiting times, we find that
the tails of the van Hove functions are only represented well by a modified model
employing a power-law jump-size distribution and stretched exponential inter-event
times. The parameters of this model confirm the heavier nature of tails in the SM as
compared to the LJ fluid. The dipolar orientational characteristics also contribute
to the power-law tail, similar to the relative displacement. Moreover, the analysis
of mean-squared displacement reveals that both fluids exhibit analogous features,
including a plateau regime at low quenching temperatures and sub-diffusive behavior
over extended simulation periods.

Incorporating magnetic inclusions adds an important feature to simple off-lattice
models that could be beneficial in addressing the challenges of modeling L\'evy
flight distributions~\cite{Metzler2000}. Magnetic colloidal glasses also serve as
analogues of dipolar glasses.  An interesting future direction is to study asymmetric
mixtures of magnetic and non-magnetic particles as a step towards colloidal spin
glasses. We anticipate that such studies may provide the nexus between structural
glasses, dipolar glasses, and spin glasses, especially because they offer direct
experimental access to particle-scale dynamics through techniques such as confocal
microscopy and scattering protocols. We hope that this work will provoke joint
experimental and theoretical investigations to provide fresh insights into these
complex systems.

\acknowledgments

The authors thank DAAD (project ID 57622781) and DST (DST/INT/DAAD/P02/2022) for
support through an Indo-German exchange scheme.

\appendix
\renewcommand{\thefigure}{S\arabic{figure}}
\setcounter{figure}{0}

\section{Description of the van Hove function from real space to the Fourier–Laplace domain}
\label{FL-space}

We begin by considering the van Hove function in real space, which is defined as
\begin{equation}  
\label{van_Hove}
G_s(\vec{\Delta}, t) = \left<\delta\left(\vec{\Delta}+\vec{r}_i(t_0)-
      \vec{r}_i(t+t_0)\right)\right>,
\end{equation}
where \(\Delta = \vec{r}(t + t_0) - \vec{r}(t_0)\). By applying the Fourier transform
to Eq.~\ref{van_Hove}, we obtain the self-intermediate scattering function, defined
as
\begin{equation}
    G_s(\vec{q}, t) = \langle e^{i\vec{q} \cdot [\vec{r}(t+t_0)-\vec{r}(t_0)]}     \rangle.
\end{equation}
Consequently, upon taking its Laplace transform, the van Hove function assumes the
following general form in Fourier–Laplace space:
\begin{equation}
    G_s(\vec{q},s) = \int_0^{\infty} dt\,e^{-st}G_s(\vec{q}, t).
\end{equation}
%


\subsection{Assumptions underlying the model}

To employ the continuous-time random walk (CTRW) formalism for the analysis of the
van Hove function, the following assumptions are introduced:
\begin{enumerate}
\item 
  The vibrational dynamics of particles confined within local cages are modeled as
  (isotropic) Gaussian-distributed displacements, defined as
  \begin{equation}
    f_\mathrm{vib}(\vec{q}) = e^{-|\vec{q}|^2 l^2/2},
  \end{equation}
  where $l^2$ denotes the mean-squared vibrational amplitude.
\item Cage-breaking events, corresponding to transitions between distinct cages, are
  characterized by a jump-length distribution. The associated characteristic function
  is given by
  \begin{equation}
    f_\mathrm{jump}(\vec{q}) = \int d^3\delta\, e^{-i\vec{q} \cdot \vec{\delta}} f_\mathrm{jump}(\vec{\delta}),
  \end{equation}
  where $f_\mathrm{jump}(\vec{\delta})$ corresponds to the jump distribution in real
  space.
\item In addition to these spatial statistics, two distinct waiting-time
  distributions are introduced. The first distribution, $\phi_1(t)$, describes the
  time elapsed until the first jump event, whereas the second, $\phi_2(t)$,
  characterizes the waiting times between all subsequent jumps. Both waiting-time
  distributions are assumed to be exponential:
  \begin{equation}
    \phi_1(t) = \frac{1}{t_1}\exp\Big{(}-\frac{t}{t_1}\Big{)},
  \end{equation}
  \begin{equation}
    \phi_2(t) = \frac{1}{t_2}\exp\Big{(}-\frac{t}{t_2}\Big{)},
  \end{equation}
  where \(t_1\) and \(t_2\) denote the corresponding characteristic waiting times.
\end{enumerate}
 
\subsection{Decomposition of particles trajectories}

The full displacement distribution is then constructed via the following steps.
\begin{enumerate}
\item The contribution from particles vibrating within cages that have not undergone a
  jump up to time $t$. The survival probability that a particle remains in its
  initial cage without performing any jumps up to time $t$ is given by
  \begin{equation}
    \tilde{\phi}_1(t) = 1 - \int_0^t \phi_1(\tau) d\tau,
  \end{equation}
  where $\phi_1(t)$ denotes the waiting-time probability density for the first
  jump. Consequently, the contribution to the self part of the van Hove function
  associated with particles that have not experienced any jumps is
  \begin{equation}
    G_s^{(0)}(\vec{q}, s) = f_\mathrm{vib}(\vec{q}) \,\frac{1-\phi_1(s)}{s},
  \end{equation}
  where $f_\mathrm{vib}(\vec{q})$ is the characteristic function corresponding to the
  vibrational motion within a cage, and $\phi_1(s)$ is the Laplace transform of
  $\phi_1(t)$.
\item The contribution from particles that have performed at least one jump.
  \begin{enumerate}
  \item \emph{First-jump contribution}: The contribution associated with the first
    jump is determined by the corresponding waiting-time probability density, whose
    Laplace transform is $\phi_1(s)$. Upon performing this first jump, the particle
    acquires a factor $f_\mathrm{jump}(\vec{q})$, which represents the characteristic
    function of the jump-length distribution, while the intra-cage vibrational
    dynamics continue to contribute a multiplicative factor
    $f_\mathrm{vib}(\vec{q})$. Thus, the combined contribution after the first jump
    is
    \begin{equation}
      \phi_1(s) \times f(\vec{q}) \times f_\mathrm{vib}(\vec{q}).
    \end{equation}
    Here $f(\vec{q}) = f_\mathrm{vib}(\vec{q})f_\mathrm{jump}(\vec{q})$.
  \item \emph{Subsequent-jump (renewal) contributions}: For all jumps following the
    first one, the dynamics are described as a renewal process. Each additional jump
    is characterized by a waiting-time distribution with Laplace transform
    $\phi_2(s)$ and the same jump characteristic function
    $f(\vec{q})$. The cumulative contribution of an arbitrary number of
    such subsequent jumps, folded with the intervening vibrational motion, yields a
    geometric series in Laplace–Fourier space:
    \begin{align}
      1 + \phi_2(s) f(\vec{q}) + [\phi_2(s) f(\vec{q})]^2 &
      + \cdots \nonumber\\
      = \frac{1}{1 - \phi_2(s) f(\vec{q})}.
    \end{align}
  \item \emph{Survival after the last jump}: The probability that no further jump occurs after the last one is given, in Laplace space, by
    \begin{equation}
      \tilde{\phi}_2(s) = \frac{1 - \phi_2(s)}{s},
    \end{equation}
    where $\phi_2(s)$ is the Laplace transform of the waiting-time distribution associated with all jumps after the first.
  \end{enumerate}
\end{enumerate}
  
Collecting these elements, the total contribution from particles that have undergone at least one jump is
\begin{equation}
  G_s^{(\geq 1)}(\vec{q}, s) = f(\vec{q}) f_\mathrm{vib}(\vec{q})
  \frac{\phi_1(s)}{s} \frac{1 - \phi_2(s)}{1 - \phi_2(s) f(\vec{q})}.
\end{equation}
Finally, summing the contributions from particles that remain vibrating in their cages without jumping and from those that participate in jump events, the total self part of the van Hove function within the CTRW formalism is given by
\begin{align}
G_s(\vec{q}, s) &= f_\mathrm{vib}(\vec{q})\frac{1-\phi_1(s)}{s} \\ \nonumber &+ f(\vec{q})f_\mathrm{vib}(\vec{q}) \frac{\phi_1(s)}{s} \frac{1-\phi_2(s)}{1-\phi_2(s) f(\vec{q})}.
\end{align}

\section{Extension of CTRM formalism for the van Hove function}
\label{M-CTRW}

We now proceed to extending the CTRW framework to the case of power-law jumps and
stretched-exponential waiting time distributions.  Within this modified or M-CTRM
framework, fitting the van Hove function requires four independent parameters. The
characteristic length scales associated with vibrational motion and jump events are
obtained from consecutive step-size data by applying a cut-off that separates
them. The remaining two parameters describe the waiting-time distribution and are
derived from the temporal statistics of the jump events. In the following, we extract
these parameters sequentially.

\subsection{Vibrational length scale}

The vibrational (Debye–Waller) factor in Fourier space is defined as
\begin{equation}
  f_{\mathrm{vib}}(q) = \exp\!\left(-\frac{q^{2} l^{2}}{2}\right),
\end{equation}
which corresponds to a Gaussian distribution of displacements in real space,
expressed as:
\begin{equation} \label{vib}
  f_{\mathrm{vib}}(\delta) = \frac{1}{\sqrt{2\pi l^2}} \exp\!\left(-\frac{\delta^2}{2l^2}\right),
\end{equation}
where $\delta$ denotes the consecutive step size in one arbitrary coordinate
direction, constrained to values smaller than or equal to the chosen threshold
$\delta_\mathrm{c} = 0.2\sigma$, and $l$ represents the characteristic vibrational
length scale, which quantifies the effective cage size associated with localized
particles.

The histogram of the step sizes $\delta \leq \delta_\mathrm{c}$ for the LJ fluid is
presented in Fig.~\ref{LJ_para}(a). The data are fitted using the Gaussian
distribution given in Eq.~\ref{vib}, from which we obtain a characteristic length
scale of $l = 0.07$. Analogously, the step-size distribution for the SM fluid,
presented in Fig.~\ref{SM_para}(a), is fitted with the same Gaussian form, yielding a
characteristic length scale of $l = 0.10$. These characteristic lengths are
interpreted as the typical amplitudes of the vibrational motion of particles confined
within cages formed by their nearest neighbors.

\subsection{Jump length scale}

The jump events are modeled by a power-law function defined as
\begin{equation} \label{jump}
f_{\mathrm{jump}}(\delta)
=\left( \delta^2+\delta_c^2 \right)^{-d/2},
\end{equation}
where $\delta > \delta_\mathrm{c}$ denotes the step size, and $d$ represents the
power-law exponent. This functional form exhibits an algebraic decay.

For the analysis of the step-size data with values exceeding the threshold
$\delta_\mathrm{c}$, we present in Fig.~\ref{LJ_para}(b) the step-size distribution
for the LJ fluid and fit it using the power-law function defined in
Eq.~\ref{jump}. From this fit, we obtain a power-law exponent of $d =
7.45$. Similarly, for the SM data, the step-size histogram is shown in
Fig.~\ref{SM_para}(b) and is fitted with the same power-law function. This procedure
yields a power-law exponent of $d = 7.68$. These two exponents effectively
characterize the slope of the corresponding distributions in the log-log
representation.


\subsection{Waiting-time distributions in Laplace space}

The Laplace-transformed waiting-time kernels are specified as
\begin{equation}
  \phi_1(s)=\frac{1}{1+(s\tau_1)^{t_1}},
\end{equation}
\begin{equation}
  \phi_2(s)=\frac{1}{1+(s\tau_2)^{t_2}},
\end{equation}
where the exponents $t_1$ and $t_2$ quantify the degree of non-exponential
waiting-time statistics, and $\tau_1$ and $\tau_2$ represent characteristic scaling
factors. The corresponding real-space (time-domain) waiting-time probability density
function is obtained via the inverse Laplace transform, defined as
\begin{equation}
  P_1(t)=\mathcal{L}^{-1}\{\phi_1(s)\}.
\end{equation}
Carrying out the inversion yields
\begin{equation} \label{waiting-time}
  P_1(t)=\frac{t^{t_1-1}}{\tau_1^{t_1}}
  E_{t_1,t_1}\!\left[-\left(\frac{t}{\tau_1}\right)^{t_1}\right],
\end{equation}
where \(E_{\alpha,\beta}(z)\) denotes the two-parameter Mittag–Leffler function,
defined by the convergent series
\begin{equation}
  E_{\alpha,\beta}(z)=
  \sum_{k=0}^{\infty}
  \frac{z^k}{\Gamma(\alpha k+\beta)},
\end{equation}
where $\alpha$ and $\beta$ are arbitrary parameters of $E_{\alpha,\beta}(z)$. Here,
$t_1$ quantifies the deviation from ordinary exponential kinetics: for $t_1 = 1$ the
function is purely exponential, while for $t_1 < 1$ it decays more slowly than an
exponential. Analogous expressions are used for $t_2$.

\begin{figure*}[tb!]
\centering
\includegraphics[width=1\textwidth]{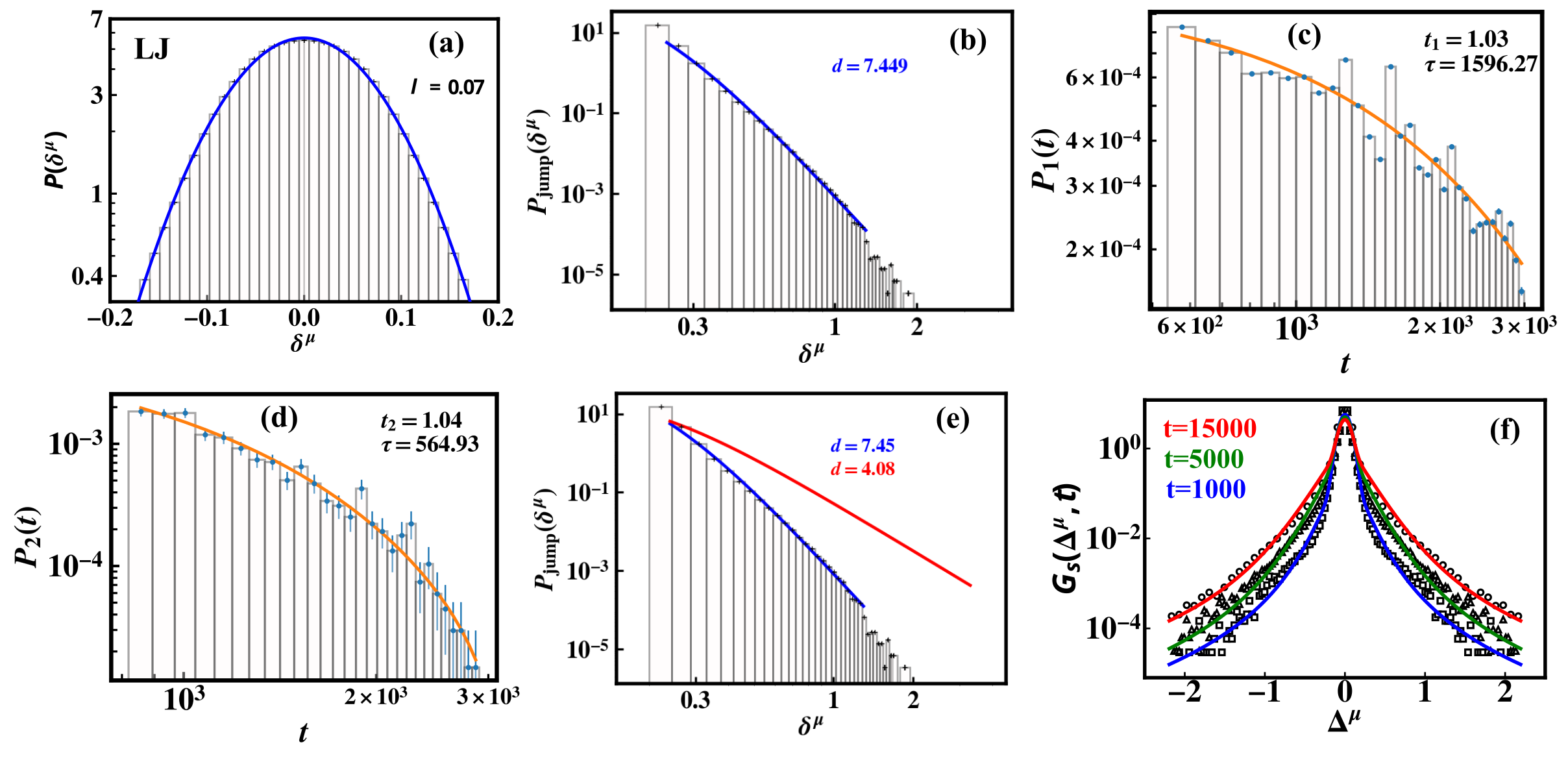} 
\caption{Fitting parameters for the van Hove function in the LJ fluid at $T =
  0.1$. (a) Characteristic vibrational length scale $l$. (b) Jump-diffusion length
  scale $d$. (c) First characteristic waiting time $t_1$. (d) Second characteristic
  waiting time $t_2$. (e) Comparison of the fitted jump length in step size $d=7.45$
  (blue) with the value used in fitting the van Hove function, $d=4.08$ (red), at
  time $t=15000$. (f) Fits of the van Hove function at $t = 1000$ (blue), $3000$
  (green), and $15000$ (red).}
\label{LJ_para}
\end{figure*}

Here, we analyze the waiting-time distribution data. Figure~\ref{LJ_para}(c) shows
the distribution of the first waiting times for the LJ fluid, which we fit using the
generalized exponential form known as the Mittag–Leffler function defined in
Eq.~\ref{waiting-time}. From this fit, we obtain a characteristic waiting time of
$t_1 = 1.03$, which is close to unity and therefore indicates that the functional
form is equivalent to an ordinary exponential decay. An analogous analysis is
performed for the subsequent waiting times, with the results for the LJ fluid shown
in Fig.~\ref{LJ_para}(d). These data are again fitted with the same Mittag–Leffler
function, yielding a consecutive waiting time of $t_2 = 1.04$. This value likewise
indicates that the tail of the distribution is well described by an approximately
exponential decay.

Similarly, for the SM fluid, Fig.~\ref{SM_para}(c) presents the distribution of the
first waiting time and its fit using the Mittag–Leffler function. The fit yields a
characteristic first waiting time of $t_1 = 0.54$. This value deviates significantly
from unity, indicating that the decay of the empirical distribution is significantly
slower than simple exponential decay. Figure~\ref{SM_para}(d) displays the
distribution of consecutive waiting times, again fitted with the Mittag–Leffler
function, resulting in $t_2 = 0.91$. This value is moderately deviated from unity and
indicates that the decay of the distribution still departs from a purely exponential
functional form.

\medskip

\subsection{Fitting of the van Hove function}

\begin{figure*}[tb!]
\centering
\includegraphics[width=1\textwidth]{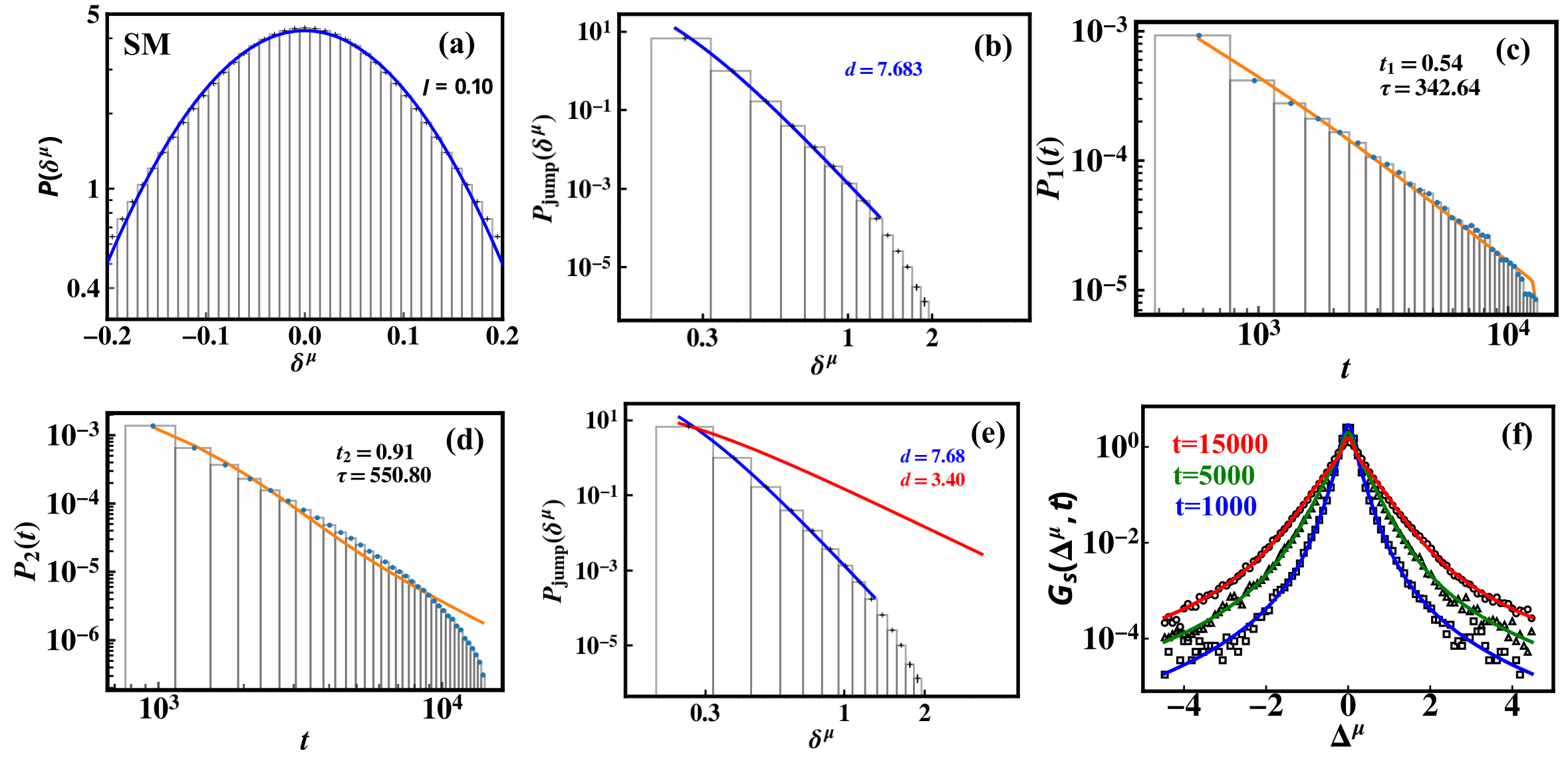} 
\caption{Fitting parameters for the van Hove function in the SM fluid at $T =
  0.4$. (a) Characteristic vibrational length scale $l$. (b) Jump-diffusion length
  scale $d$. (c) First characteristic waiting time $t_1$. (d) Second characteristic
  waiting time $t_2$. (e) Comparison of the fitted jump length $d=7.68$ (blue) with
  the applied jump length $d=3.40$ (red) in the van Hove function. (f) Fits of the
  van Hove function at $t = 1000$, $4000$, and $15000$.}
\label{SM_para}
\end{figure*}

We have determined the fitting parameters using the M-CTRW formalism for both LJ and
SM fluids. Figure~\ref{LJ_para}(f) presents the fits to the self part of the van Hove
function for the LJ fluid at times $t = 1000$, $5000$, and $15000$~\footnote{To
  perform the fits, we use a numerical implementation of the inverse Fourier-Laplace
  transform, adapting the fit parameters $t_1$, $t_2$, $\tau_1$, $\tau_2$, $l$,
  $\delta_c$ and $d$ to achieve the best fit.}. In all cases, the parameters perform
satisfactorily; however, accurate fits require different values of $d$, as shown in
Fig.~\ref{LJ_para}(e). This variation arises from the presence of a larger number of
mobile surface particles, whose correlated vibrational motion and jumps enhance the
contribution to the tails of the distribution. Overall, we find that the distinction
between vibrations and jumps is not as clear cut as initially assumed and,
additionally, there exist correlations between subsequent jump events.
 
Similarly, we have performed the fitting of the van Hove function for the SM fluid at
times $t = 1000$, $5000$, and $15000$, as shown in Fig.~\ref{SM_para}(f). In this
case as well, all parameters perform well except for $d$, which requires a smaller
value, as can be seen by comparison with the corresponding values in
Fig.~\ref{SM_para}(e).

\bibliography{SM_fluid}

\end{document}